\documentclass[12pt,a4paper]{iopart}
\usepackage[T1]{fontenc}
\usepackage[latin9]{inputenc}

\expandafter\let\csname equation*\endcsname\relax
\expandafter\let\csname endequation*\endcsname\relax

\usepackage{cite} % sort&compress references
\usepackage{amsmath}
\usepackage{amssymb}
\usepackage{graphicx}
\usepackage{float}
\usepackage{esint}
\usepackage{mathrsfs}
\usepackage{bm}
\usepackage{bbm}
\usepackage{bbold}
\usepackage[all]{xy}
\usepackage{tikz} 
\usepackage{pgfplots}
\usepackage{enumitem}
\usepackage{fancyhdr}
\usepackage{color}
\usepackage[breaklinks]{hyperref}
\hypersetup{colorlinks,citecolor=blue,linkcolor=blue,urlcolor=black}
\def\com{\color{magenta}}
\def\cob{\color{blue}}

\newcommand{\Eq}[1]{(\ref{#1})}
\newcommand{\au}[2]{#2 #1}
\newcommand{\oarX}[1]{\href{http://arxiv.org/abs/#1}{(arXiv:{\com #1})}}
\newcommand{\arX}[1]{\href{http://arxiv.org/abs/#1}{(arXiv:{\com #1})}}
\newcommand{\doin}[6]{\href{http://dx.doi.org/#1}{\cob \textit{#2} #3 \textbf{#4} #5}}
\newcommand{\doit}[5]{\href{http://dx.doi.org/#1}{\cob #2 \textbf{#3} (#5) #4}}

\newcommand{\doinn}[5]{\href{http://dx.doi.org/#1}{\cob \textit{#2} \textbf{#3} #4}}

\newcommand{\doij}[5]{\href{http://dx.doi.org/#1}{\cob \textit{#2} #3(#5)#4}}
\newcommand{\tia}[1]{#1}
\newcommand{\books}[4]{#4 \emph{#1} (#3: #2)}

\numberwithin{equation}{section}

\makeatletter

\def\ds{d_{\rm S}}
\newcommand{\ra}{\rightarrow}

\newcommand{\In}{\subset}

\newcommand{\C}{\mathbb C}
\newcommand{\R}{\mathbb R}
\newcommand{\Z}{\mathbb Z}
\newcommand{\N}{\mathbb N}

\def\d{\mbox d}

\makeatother

\begin{document}

\begin{flushright} \doit{10.1088/0264-9381/31/13/135014}{Class.\ Quantum Grav.}{31}{135014}{2014} \hfill  \arX{1311.3340}\\ AEI-2013-196\end{flushright}

\date{November 13, 2013}

\title{Spectral dimension of quantum geometries}

\author{Gianluca Calcagni,}
\address{Instituto de Estructura de la Materia -- CSIC, calle Serrano 121, E-28006 Madrid, Spain}
\author{Daniele Oriti, Johannes Th\"urigen}
\address{Max Planck Institute for Gravitational Physics (Albert Einstein Institute)\\
Am M\"uhlenberg 1, D-14476 Potsdam, Germany}
\eads{\mailto{calcagni@iem.cfmac.csic.es}, \mailto{doriti@aei.mpg.de}, \mailto{johannes.thuerigen@aei.mpg.de}}

\begin{abstract}
The spectral dimension is an indicator of geometry and topology of spacetime and a tool to compare the description of quantum
geometry in various approaches to quantum gravity. This is possible
because it can be defined not only on smooth geometries but also on
discrete (e.g., simplicial) ones. In this paper, we consider the spectral dimension
of quantum states of spatial geometry defined on combinatorial complexes endowed with additional algebraic data: the kinematical quantum states of
loop quantum gravity (LQG). Preliminarily, the effects of topology and discreteness of classical discrete geometries are studied in a systematic manner.
We look for states reproducing the spectral dimension of a classical space in the appropriate regime. We also test the hypothesis that in LQG, as in other approaches, there is a scale dependence of the spectral dimension, which runs from the topological dimension at large scales to a smaller one at short distances. While our results do not give any strong support to this hypothesis, we can however pinpoint when the topological dimension is reproduced by LQG quantum states. Overall, by exploring the interplay of combinatorial, topological and geometrical effects, and by considering various kinds of quantum states such as coherent
states and their superpositions, we find that the spectral dimension of discrete quantum
geometries is more sensitive to the underlying combinatorial structures than to the details of the additional data associated with them.
\end{abstract}

\hspace{1.5cm} {\footnotesize Keywords: spectral dimension, quantum gravity, discrete geometry, loop quantum gravity}

\hspace{1.5cm} {\footnotesize  gravity}

% 02.40.Sf	Manifolds and cell complexes
% 04.60.-m	Quantum gravity
% 04.60.Pp	Loop quantum gravity, quantum geometry, spin foams


\maketitle

\pagestyle{fancy}
\renewcommand{\sectionmark}[1]{\markright{\thesection\ #1}}

\rhead{\fancyplain{}{}} % or predefined in second bracket "Spectral dimension of quantum geometries"
\lhead{\fancyplain{}{\rightmark }}

\tableofcontents

%%%%%%%%%%%%%%%%%%%%%%%%%%%%%%%%%%%%%%%%%%%%%%%%%%%%%%%%%%%%%%%%%%%%%%%%%%%%%%%%%%%%%%%%%%%%%%%%

\section{Introduction}

The spectral dimension $\ds$ of quantum geometries has been a rather hot topic in recent quantum gravity research. $\ds$ is one of the indicators of topological and geometric properties of quantum spacetime as described by different approaches, as well as a way to check that an effective semi-classical spacetime is obtained in appropriate sectors of the theory.
One of the main goals of any such analysis is to prove that, among all the states and histories that appear in the theory (most of which far from describing any smooth spacetime geometry), configurations which do so are either dominant or the approximate result of averaging over the others. 
A further goal is to study whether the effective dimension of spacetime remains the same at different scales or if it is subject to a dynamical reduction in the ultraviolet regime \cite{tHo93,Car09,fra1}.

In dynamical triangulations \cite{Ambjorn:2010kv,Ambjorn:2012vc}, the spectral dimension has been important as one of the few observables available to classify phases of spacetime ensembles. 
In particular, in {\it causal dynamical triangulations} (CDT) in four dimensions it was found that, while very close to the spacetime topological dimension $d+1=4$ at large scales, at least in the \lq geometric phase\rq, it takes smaller values at short distances, approaching the value $\ds\simeq 2$ in the ultraviolet limit \cite{Ambjorn:2005fj,Ambjorn:2005fh}. Similar results are obtained in three dimensions \cite{Benedetti:2009bi}. Two is, of course, an interesting value for the effective dimension in the ultraviolet, because gravity is perturbatively renormalizable in two dimensions. In fact, also in the asymptotic safety scenario \cite{Niedermaier:2006up,Reuter:2012jx} one can find a reduction of the spectral dimension from the topological dimension $d+1$ to half of it under the renormalization group flow using analytical methods \cite{Lauscher:2005kn,Reuter:2013ji,CES}. Other formalisms manifest similar behaviour. Ho\v{r}ava--Lifshitz gravity has a built-in dynamical dimensional reduction, due to the defining anisotropic scaling \cite{CES,Horava:2009ho}. 
Modified dispersion relations in non-commutative spacetimes \cite{Benedetti:2009fo,Alesci:2011wi}, super-renormalizable non-local gravity \cite{Mod1}, black-hole effective physics \cite{AC1} and other scenarios (including string theory \cite{CM1}) give rise to a dimensional flow as well.

Apart from the intrinsic interest in this effect, its occurrence in approaches to quantum gravity which are otherwise hard to compare has raised hope that the spectral dimension might help to relate such approaches to one another, or at least to confront their results \cite{frc4}. Although the spectral dimension is a rather coarse tool in comparison with more refined geometric indicators \cite{CES,frc7}, its determination in full-fledged quantum geometries is still a technically challenging problem.

All the above considerations gave good reasons to investigate the spectral dimension also in loop quantum gravity (LQG) \cite{Rovelli:2004wb,Thiemann:2007wt,Perez:2013uz, LQG-GFT, Oriti:2012wt}. Using the spectrum of the area operator in LQG to guess an effective dispersion relation for propagating particles, some evidence was found for a dimensional flow of kinematical LQG states \cite{Modesto:2009bc} and their spacetime spin-foam dynamics \cite{Caravelli:2009td,Magliaro:2009wa}. In the present work, we go beyond the arguments used in previous analyses and investigate the spectral dimension of LQG states in detail, at least at the level of kinematical states. In particular, we will study the spectral dimension for states defined on large graphs or complexes.

LQG differs from all the above-mentioned approaches in that the degrees of freedom of the theory are discrete geometric data, i.e., a combination of holonomies, fluxes or Lorentz group representations \cite{Rovelli:2004wb,flux} associated with combinatorial structures such as graphs and cellular complexes. This combination poses new challenges.
So far, the spectral dimension has been investigated in more or less traditional smooth settings, in terms of either modified differential structures (multi-scale spacetimes \cite{frc7}) or modified dispersion relations on a smooth geometric manifold (asymptotic safety, where anomalous scaling is a general feature at the non-Gaussian fixed point; non-commutative spacetimes, where modified dispersion relations are a consequence of a deformed Poincar\'e symmetry; Ho\v{r}ava--Lifshitz gravity, where the Laplacian is multi-scale by construction; and so on). Also previous work on the LQG case \cite{Modesto:2009bc} has been limited to an approximation in which the scaling of the Laplacian is first extracted from the area spectrum and then effectively treated as continuous. In other approaches, the setting is purely combinatorial: such is the case of CDT (the dynamical degrees of freedom are equilateral triangulations) or graph models \cite{Gia13}. 

Therefore, the first important step is a definition of the spectral dimension on the type of discrete geometries LQG is built on\footnote{There are approaches to quantum gravity where spacetime structures are even more peculiar, but where notions of dimension nevertheless still can be defined: In causal sets dimension estimators such as the so called Myrheim--Meyer or the midpoint dimension are used to determine, via the local Minkowski structure, the dimension of the curved spacetimes which causal sets can be faithfully embedded into \cite{Reid:2003jc}.}.
%%%
This preliminary work has already been carried out in \cite{COT1,Thuerigen:2013vc}, where we have generalized discrete exterior calculus \cite{Desbrun:2005ug} to be applicable to such discrete geometries, and we have given explicit expressions for the Laplacian (the basic geometric operator the spectral dimension depends on) in terms of the variables relevant for LQG.

Unless stated otherwise, from now on by `spectral dimension' we always mean the one \emph{of spatial (not spacetime) geometries}. Also, by \emph{quantum spectral dimension} of a quantum state of geometry we indicate the spatial spectral dimension as derived from the expectation value of the heat kernel trace on the quantum geometry. In the usual interpretation, this heat kernel is the solution of the diffusion equation for a test pointwise particle propagating on the quantum geometry, where the differential variable is an abstract `diffusion time' $\tau$. One can employ a different and perhaps more convincing interpretation \cite{CM1,CMN}, where the diffusion equation is a running equation with respect to the resolution $1/\sqrt{\tau}$ at which geometry is probed. We will explain why, in the case of discrete quantum gravity, for a well-defined notion of quantum spectral dimension the quantum expectation value has to be taken at the level of the heat kernel, and not directly of the Laplacian.

The combination of combinatorial discreteness and additional (pre-)geometric data obviously must result in an interplay of their respective effects. To the best of our knowledge, the latter have never been classified in the literature. A major aim of this paper is to understand this interplay in the case of kinematical LQG states.
%Which features come from the underlying discreteness, which are due to the geometric data and how do they interact?
We first study systematically the topological and geometric effects in the spectral dimension of smooth spheres and tori as well as the discreteness effects of combinatorial complexes with geometric realizations as (hyper-)cubulations or triangulations of these.
This groundwork having been done, we can then compare the spectral dimension of quantum states with that of the discrete geometries they are semi-classical approximations or superpositions of. As a general tendency, we find that the effect of the underlying combinatorics dominates and that a relatively large size of the base complexes is needed for the spectral dimension to behave as a spatial dimension at all.

To construct explicit LQG states on large complexes and calculate the corresponding expectation values of global observables is a major and seldom accepted challenge. In the case of the spectral dimension, however, this becomes inevitable. As a line of attack, we set up a way to numerically construct large abstract simplicial complexes and define geometries thereon. We follow \cite{Bell:2011wu} to derive their combinatorial properties needed for studying the action of the Laplacian operator.

%\

While the preparatory work on the classical geometries is done in arbitrary dimensions, we restrict the quantum analysis to kinematical states of (2+1)-dimensional Euclidean LQG. There are two reasons for this restriction. 

The first is merely technical. The number of degrees of freedom for given assignments of algebraic data grows with the combinatorics of the complex as $n \sim p^d$, where $d$ and $p$ are, respectively, the topological dimension and the size of the combinatorial complex. The Laplacian operator defined thereon is an $n\times n$ matrix, which makes the computational effort more severe as $d$ increases. Thus, for a given $p$ one can calculate much larger $d=2$ discrete geometries than in $d=3$. On the other hand, a sufficient combinatorial size turns out to be crucial for the physical interpretation of the spectral dimension as a notion of positive-definite dimension somewhere close to $d$. This limitation could be easily overcome by using a more powerful computational environment for numerical analysis.

The second reason lies in the structure of quantum gravity itself. In 2+1 dimensions, the spin-network basis of LQG simultaneously diagonalizes all edge-length operators, permitting a straightforward definition of the Laplacian and a direct identification of deviations of the quantum spectral dimension from its classical counterpart as quantum corrections. In 3+1 dimensions, one would have to deal further with effects due to the non-commutativity of the full set of geometric operators needed to define the Laplacian, as well as with the role of non-geometric configurations which we know are present in the standard $SU(2)$-based Hilbert space of the theory. Its construction thus becomes much more involved and it is much harder to be implemented efficiently for numerical calculations. While these challenges do not pose any obstacle in principle, we prefer to concentrate here on the more straightforward 2+1 case. Once the general influence of quantum geometric effects (discreteness, quantum fluctuations, and so on) on the spectral dimension is understood, one can move on to study the additional complications that are present in the four-dimensional case.

This paper is quite explorative in nature. There are no constraints on the kind of combinatorial manifolds LQG states could be defined on. Thus, the only guidance may come from classical geometries one might want to approximate semi-classically by coherent states. Furthermore, there is no unique way to parametrize the `quantum-ness' of states. In fact, to search for genuine quantum effects of LQG states, it is not enough to pick one state and look only at the value of its spectral dimension at small scales, as is usually done when some kind of dimensional flow is expected. Instead, we compute the spectral dimension for those states we consider as `more quantum'. These are coherent spin-network states peaked at smaller representation labels and with larger spread and, more generally, states highly randomized in the intrinsic geometry or superposed with respect to various geometric degrees of freedom. Moreover, in order to explore quantum effects of the combinatorial structure and to compare with CDT, we consider superpositions of these structures as well.

\subsection{Outline}

In section \ref{sec:dsRiem}, the spectral dimension is introduced by the common definition on Riemannian manifolds. We present then a definition on discrete geometries. This is the basis whereupon to define the quantum spectral dimension in general, and explicitly in the case of (2+1)-dimensional LQG, which we review.

In section \ref{sec:Topo}, geometric and topological features are discussed, both analytically and numerically, in particular for the simple examples of spheres and tori.
In section \ref{sec:dsDiscrete}, we study the features of the spectral dimension on discrete structures, exemplified through various cases.
These preliminary considerations are carried out also in higher dimensions. This part of the paper classifies phenomena which can appear in all contexts where a discrete spacetime structure is employed, both in analytic and numerical applications; hence, it may be of interest also for the general reader not specialized in quantum gravity.

In section \ref{sec:dsKinematics}, the expectation value of
kinematical quantum space states in 2+1 LQG is analyzed. Here the focus
is on investigating the relevance, for the spectral dimension, of the geometric data additional
to the geometry coming from purely combinatorial structures. 
The spectral dimension of semi-classical states is compared with the classical one to identify quantum corrections, which turn out to be rather small. 
Furthermore, superpositions of coherent states peaked at different spins will be discussed; the spectral dimension of these configurations will be found via a certain averaging procedure.

Complementary to these superpositions in geometry, in section \ref{sec:sumDis} we analyze the spectral dimension of superpositions of states on two classes of torus triangulations. It turns out that the quantum spectral dimension in LQG is more sensitive to the underlying graph structure of states than to the associated geometric data.

%%%%%%%%%%%%%%%%%%%%%%%%%%%%%%%%%%%%%%%%%%%%%%%%%%%%%%%%%%%%%%%%%%%%%%%%%%%%%%%%%%%%%%%%%%%%%%%%

\section{Definition of discrete and quantum spectral dimension}

In this section, we start by introducing the concept of spectral dimension in the usual context of  Riemannian manifolds. We show how this definition can be transferred to discrete pseudo-manifolds, i.e., abstract finite simplicial complexes with manifold properties. It is straightforward then to define the quantum spectral dimension in terms of expectation values on quantum states. For kinematical states of LQG in 2+1 dimensions, we argue that this expression is well defined under rather mild conditions.

\subsection{Spectral dimension of Riemannian geometries\label{sec:dsRiem}}

The definition of the spectral dimension is related to the scaling
of the propagator of a massless test particle field (or, equivalently, to a diffusion process on the spatial manifold). $\ds$ is defined via the heat kernel
$K(x,x';\tau)$, which describes the diffusion of the particle in fictitious
`time' $\tau$ from point $x'$ to $x$. $K$ is the solution of the diffusion equation
\begin{equation}
\partial_{\tau}K(x,x';\tau)-\Delta_{x}K(x,x';\tau)=0\,,\label{eq:HeatKerEq}%\qquad K(x,y;0)=\delta(x-y)\,,
\end{equation}
 where $\Delta$ is the Laplacian with respect to the underlying space. Actually, $\tau$ has dimension of a squared length since we set the diffusion parameter in front of the Laplacian to 1. Alternatively, equation \Eq{eq:HeatKerEq} can be regarded as a running equation establishing how much we can localize the point particle, placed at some point $x'$, if we probe the geometry with resolution $1/\sqrt{\tau}$ \cite{CMN}. The length scale $\sqrt{\tau}$ represents the minimal detectable separation between points. While in the diffusion interpretation $K$ is the probability to find the particle at point $x$ after diffusing for some time $\tau$ from point $x'$, in the `resolution' interpretation $K$ is the probability to see the particle in a neighborhood of point $x$ of size $\sim\sqrt{\tau}$ if the geometry is probed with resolution $1/\sqrt{\tau}$. For infinite resolution ($\tau=0$) and a delta initial condition, the particle is found where it was actually placed (at $x'$), while the smaller the resolution (i.e., larger diffusion time) the wider the region where it can be seen. There is no practical difference between the two interpretations but the latter is more suitable in diffeomorphism-invariant theories where the notion of diffusion time makes no sense (while a Lorentz-invariant scale does).

A formal solution to the problem is provided by
\begin{equation}
K(x,x';\tau)=\left\langle x'|\exp(\tau\Delta)|x\right\rangle\,, \label{eq:Kformal}
\end{equation}
where the initial condition is incorporated as the definition of the
inner product $\left\langle \cdot|\cdot\right\rangle $ of particle
states. Usually, the particle is taken to be concentrated at $\tau=0$
in the point $x'$, but it could start with other distributions. 

In the usual context of a Riemannian $d$-manifold $(M,g)$, the Beltrami--Laplace
operator is (on 0-forms the connection Laplacian and Hodge Laplacian agree) 
\begin{align}
\Delta^{g} & =\mbox{Tr}\nabla^{2}=g^{\mu\nu}\nabla_{\mu}\nabla_{\nu}=g^{\mu\nu}\left(\partial_{\mu}\partial_{\nu}-\Gamma_{\mu\nu}^{\rho}\partial_{\rho}\right)=\frac{1}{\sqrt{g}}\partial_{\mu}\sqrt{g}g^{\mu\nu}\partial_{\nu}\,.
\label{eq:LaplaceBeltrami}
\end{align}
With the initial condition
\begin{equation}
K(x,x';0^{+}) = \langle x' | x \rangle = \frac{1}{\sqrt{g}}\delta(x-x')\,,
\end{equation}
the heat kernel is
\begin{equation}
K(x,x';\tau)=\frac{\rme^{-\frac{D^{2}(x,x')}{4\tau}}}{(4\pi\tau){}^{\frac{d}{2}}},
\label{eq:Kheatg}
\end{equation}
with $D(x,x')$ the geodesic distance. On $(M,g)$ there is a well-defined (the so called Seeley--DeWitt) expansion of the heat kernel around the solution
% (\ref{eq:Kheatg})
on a flat manifold $K_{0}(x,x';\tau)$ with the metric distance being the Euclidean norm $D(x,x')=\sqrt{|x-x'|^2}$, 
\begin{equation}
K(x,x';\tau)=K_{0}(x,x';\tau)\underset{n=0}{\overset{\infty}{\sum}}b_{n}(x,x')\,\tau^{n},
\end{equation}
where the $b_{n}(x,x')$ are computable in terms of geometric invariants \cite{Vassilevich:fw}.

The spectral dimension probed by the particle can now be defined as the scaling of the trace of the heat kernel
\begin{equation}\label{eq:heattraceUnnormed}
P(\tau) = \Tr K(x,x';\tau) = \int_M \d^d x\, \sqrt{g}\, K(x,x;\tau)\,,
\end{equation}
called return probability. 
%In terms of the spectrum of the Laplacian given by eigenvalues $\lambda_{j}$ this is \begin{equation} P(\tau)=\frac{1}{V}\underset{j}{\sum}e^{-\lambda_{j}\tau}.\end{equation}

While in the flat case $P(\tau) = (4\pi\tau)^{-d/2} V$, where $V=\int_M \d^d x$, the general heat-kernel expansion yields a power series 
\begin{equation}
P(\tau)=\frac{V}{(4\pi\tau)^{\frac{d}{2}}}\overset{\infty}{\underset{n=0}{\sum}}a_{n}\tau^{n}
\end{equation}
where $V=\int_M \d^d x\,\sqrt{g}$ and the coefficients $a_{n}=\mbox{Tr}\,b_{n}$ consist of geometric invariants averaged over the manifold.

The spectral dimension $\ds$ can be implicitly defined as the scaling of the heat trace in some limit of $\tau$,
\begin{equation}
P(\tau)\simeq \tau^{-\frac{\ds}{2}},
\label{eq:dSimplicit}
\end{equation}
such that $\ds=d$ for the flat case, while in general there are corrections related to global properties described by the geometric invariants $a_{n}$. 
The implicit definition (\ref{eq:dSimplicit}) can be generalized as
\begin{equation}
\ds(\tau):=-2\frac{\partial\ln P(\tau)}{\partial\ln\tau}.\label{eq:dSexplicit}
\end{equation}
By definition (via heat trace $P$, heat kernel $K$ and Laplace operator $\Delta$) the spectral dimension is a functional of the geometry $\left[g\right]$ on
the given manifold $M$.

In most of the literature on quantum gravity, equation \Eq{eq:HeatKerEq} is taken on Euclideanized spacetime but, as stressed in the introduction, we will strictly confine our analysis to \emph{spatial} geometries. Consequently, $d=3$ is the case that fits large scale, semiclassical observations.

\subsection{Spectral dimension of discrete geometries}

From a quantum gravity perspective, the spectral dimension is particularly
interesting because it can be rather easily extended to discrete
geometries as well. Discrete geometries constitute the degrees of
freedom of LQG, spin foams and group field theory (and of CDT in the purely combinatorial case, that is when the graph distance is chosen to specify completely the geometry).

Let us briefly recapitulate the framework of discrete calculus needed for the definition of the spectral dimension on discrete geometries provided in \cite{COT1}. 
In the most general context we deal with combinatorial complexes, in particular abstract finite simplicial $d$-pseudo-manifolds ${\cal K}$,
with additional (pre-)geometric data, i.e., volume variables associated
with simplices or cells. If ${\cal K}$ has a geometric realization $|{\cal K}|$ on some metric space, it can be thought of as the triangulation of a smooth geometry $(M,g)$. But this needs not be the case.

In this setting, there is a definition of the Laplace operator in terms
of discrete exterior calculus, giving a precise meaning to the formal
expression (\ref{eq:Kformal}). To this end,
we take the diffusing scalar test particle field $\phi$ as a field living on
the $d$-simplices $\sigma$, or equivalently on the vertices of the
combinatorially dual complex. Then, the action of the Laplacian on
$\phi$ evaluated at some simplex $\sigma$ takes the form 
\begin{equation}
-\left(\Delta\phi\right)_{\sigma}=\frac{1}{V_{\sigma}}\underset{\sigma'\sim\sigma}{\sum}\frac{V_{\sigma\cap\sigma'}}{l_{\sigma\cap\sigma'}}\left(\phi_{\sigma}-\phi_{\sigma'}\right)\,,
\label{eq:disLap}
\end{equation}
where the sum runs over neighboring simplices and $V_{\sigma}$, $ $$V_{\sigma\cap\sigma'}$
and $l_{\sigma\cap\sigma'}$ are, respectively, the $d$-volume of $\sigma$, the
$(d-1)$-volume of the common face and a dual length through that
face.

%%% Expansion of Laplacian discussion
By definition, this Laplacian has most of the properties of the continuum Laplacian. That is, it obeys the null condition, it is self-adjoint (which relates to the symmetrizability of its matrix) and it is local, i.e., the action on a given position depends only on neighboring positions. 
In \cite{COT1} we have also argued for the convergence to the continuum Laplacian under refinement of triangulations.
Nevertheless, there are properties which are further dependent on the definition of the volumes. We have argued that a definition in terms of a barycentric dual interpretation of the volumes should be preferred to a  circumcentric one in the field theoretic context because only the former preserves positivity of the Laplacian. For this reason, we will adhere to this convention throughout this paper.

If we further assume that the volumes associated with the finite simplicial complex $\mathcal{K}$ are finite and non-degenerate, in particular non-vanishing, the matrix elements of the Laplacian are finite and well defined. The Laplacian is then diagonalized by its eigenfunctions $e^{\lambda}$ corresponding to eigenvalues $\lambda$,
\begin{equation}
\left(-\Delta e^{\lambda}\right)_{\sigma} = \lambda e_{\sigma}^{\lambda},
\end{equation}
and these form a complete orthonormal basis defining momentum space. Using the transformation to momentum space 
%%%
the heat trace of $\phi$ turns our to be \cite{COT1}
\begin{equation}
P(\tau)=\mbox{Tr}\left\langle \sigma'|\rme^{\tau\Delta}|\sigma\right\rangle =\underset{\lambda\in {\rm Spec}(\Delta)}{\sum}\rme^{-\tau\lambda},
\end{equation}
% where by convention the eigenvalues $\lambda$ are all positive semi-definite on compact geometries. 
Since $\Delta$ is symmetrizable, for real geometric volume coefficients in equation (\ref{eq:disLap}) the spectrum Spec and thus the heat trace are real valued.

In the trivial case of constant volumes over the complex, e.g., equilateral
triangulations, the Laplacian is just proportional to the combinatorial
graph Laplacian of the dual 1-skeleton of the complex ${\cal K}$.

\subsection{Spectral dimension of quantum states of geometry}

In a quantum theory of geometry,
the metric or other classically equivalent variables are promoted to
operators. The spectral dimension as an observable is a functional
of these operators and we are now interested in its expectation value
for a given state $\psi$ of space geometry (using the heat trace (\ref{eq:heattraceUnnormed})). One possibility is to take the heat trace as the primary observable to be quantized. Formally,
\begin{equation}
\ds^{\psi}(\tau):=-2\frac{\partial\ln\left\langle \widehat{P(\tau)}\right\rangle_{\psi}}{\partial\ln\tau}=-2\frac{\partial\ln\left\langle \mbox{Tr}\widehat{K(x,x';\tau)}\right\rangle _{\psi}}{\partial\ln\tau}\label{eq:dSquantum}.
\end{equation}
Suppose the quantum Laplacian $\widehat{\Delta}$, operator on the
Hilbert space of geometry states, is diagonalizable with orthonormal
eigenbasis $|s\rangle$ (later we will take the states $|s\rangle$ to be
the spin-network basis of the kinematical LQG Hilbert space). Then every state can be expanded
in this basis, $|\psi\rangle=\underset{s}{\sum}\psi(s)|s\rangle$,
and the heat trace simplifies:
\begin{align}
\left\langle \widehat{P(\tau)}\right\rangle _{\psi} & 
= \langle \psi|\mbox{Tr}\,\rme^{\tau\widehat{\Delta}}|\psi\rangle =\underset{s}{\sum}\left|\psi(s)\right|^{2}\langle s|\mbox{Tr}\,\rme^{\tau\widehat{\Delta}}|s\rangle \\
 & =\underset{s}{\sum}\left|\psi(s)\right|^{2}\mbox{Tr}\,\rme^{\tau\langle s|\widehat{\Delta}|s\rangle }.
\end{align}
An explicit expression of the spectral dimension (\ref{eq:dSquantum}) is thus
\begin{equation}
\ds^{\psi}=-2\tau\frac{\underset{s}{\sum}\left|\psi(s)\right|^{2}\mbox{Tr}\langle s|\widehat{\Delta}|s\rangle \rme^{\tau\langle s|\widehat{\Delta}|s\rangle }}{\underset{s}{\sum}\left|\psi(s)\right|^{2}\mbox{Tr}\,\rme^{\tau\langle s|\widehat{\Delta}|s\rangle}}.
\label{eq:qds}
\end{equation}
In the following, we will drop the superscript $\psi$ in the spectral dimension.

Alternatively, one might consider the Laplacian instead of the heat kernel as the fundamental
observable for the spectral dimension. This would give a different result with a product instead of the sum over the eigenbasis. From 
\begin{equation}
\partial_{\tau}K(x,x';\tau)=\langle \widehat{\Delta}\rangle _{\psi}K(x,x';\tau)\,,
\end{equation}
there follows
\begin{equation}
P^{\psi}(\tau)
\propto\mbox{Tr}\,\rme^{\tau\underset{s}{\sum}\left|\psi(s)\right|^{2}\langle s|\widehat{\Delta}|s\rangle }=\mbox{Tr}\underset{s}{\prod}\rme^{\tau\left|\psi(s)\right|^{2}\langle s|\widehat{\Delta}|s\rangle}.
\end{equation}
On states $s$ with discrete support, e.g., on the simplicial complexes
considered here, this formal expression makes sense only if all states
live on the same discrete structure. Otherwise, the product of
$\rme^{\langle s|\widehat{\Delta}|s \rangle}$ and $\rme^{\langle s'|\widehat{\Delta}|s'\rangle}$
is not defined, since it would amount to a sum of matrices of different size in the exponent. 
In theories of quantum gravity with discrete geometries as degrees of freedom, such as LQG, the full Hilbert space usually contains states (including superpositions) on various distinct combinatorial structures. For this reason, the possibility of defining the quantum spectral dimension as derived from the expectation value of the Laplacian has to be excluded. 

A similar issue arises when taking the expectation value at the level of the heat kernel $K$, since it is a function of positions on the (discrete) manifold. It can only be avoided by tracing over the manifold.
Therefore, we will take the expression in terms of the expectation value of the heat trace, equation (\ref{eq:dSquantum}), as the proper definition of quantum spectral dimension.
This is the same as used in CDT \cite{Ambjorn:2005fj,Ambjorn:2005fh,Benedetti:2009bi}, 
while in the smooth approaches where this issue does not arise
\cite{Lauscher:2005kn,Reuter:2013ji,CES,Horava:2009ho,Benedetti:2009fo,Modesto:2009bc,Magliaro:2009wa,Caravelli:2009td}
the expectation value is taken at the level of the Laplacian.

As a side remark, notice that using the `normed' heat trace 
\begin{equation}
\tilde{P}(\tau):=\frac{1}{V}P(\tau)=\frac{1}{V}\mbox{Tr}K(x,x';\tau)\label{eq:heattraceNormed}
\end{equation}
instead of $P(\tau)$ would lead to a classically equivalent definition of the spectral dimension (\ref{eq:dSexplicit}), as its scaling is independent of proportionality factors. This definition is very often used for effective continuous spacetimes in quantum gravity, as it gets rid of an infinite volume prefactor in $P$. However, at the quantum level and in the case of superposition states, there is a notable difference. For the volume-normed heat trace (\ref{eq:heattraceNormed}),
\begin{equation}
\left\langle \widehat{\tilde P(\tau)}\right\rangle_{\psi}=\underset{s}{\sum}\left|\psi(s)\right|^{2}\langle s|\widehat{V^{-1}}|s\rangle \mbox{Tr}\,\rme^{\tau\langle s|\widehat{\Delta}|s\rangle}\,,
%\nonumber
\end{equation}
states are weighted by an additional inverse-volume factor $\left\langle s|V^{-1}|s\right\rangle$. Since for quantum geometries this factor might have singular behaviour (for instance, on states of degenerate geometry where the volume operator is not densely defined and has 0 as an eigenvalue), we prefer equation \Eq{eq:heattraceUnnormed} to define the quantum spectral dimension.

\subsection{Heat trace operator on 2+1 kinematical LQG states
\label{sec:defLQG}}

LQG states are functions
of spatial geometric variables (holonomies, flux variables or representation
labels) smeared over graphs $\Gamma$ which we restrict to be dual
1-skeletons of a discrete pseudo-manifolds ${\cal K}$, to be able to use
%, i.e. $\Gamma=(\star K)_{1}$,
 the discrete Laplacian (\ref{eq:disLap}).
Furthermore, as anticipated and motivated in the introduction, we will restrict here to (2+1)-dimensional LQG. 
%This is not only technically simpler but allows in particular to directly attack the question of the impact of quantum fluctuations on the spectral dimension since all states are geometric. In 3+1 dimensions one would further have to deal with the issue of geometricity. This would demand in particular either a generalization of the simplicial Laplacian eq. \ref{eq:disLap} to the nonsimplicial area and volume variables in the spin representation or to work with its noncommutative flux representation as given in \cite{COT1} .

Now, the elements of $\mathcal{H}_{\rm kin} = \bigoplus_\Gamma \mathcal{H}_\Gamma$ are functions of the holonomies
on the links of the dual graph $\Gamma=(\star {\cal K})_{1}$ of a simplicial complex ${\cal K}$
% fundamentally discrete LQG here!
corresponding to a $d=2$ spatial slice\footnote{Labeling the spaces $\mathcal{H}_\Gamma$ by the dual graphs $\Gamma$ instead of the corresponding simplicial pseudo-manifolds is allowed because their relation is one-to-one. This can be easily seen since dual 1-skeletons of simplicial pseudo-manifolds are colorable and colorable graphs are dual to simplicial pseudo-manifolds \cite{Gurau:2010iu}.}.
Since $\Gamma$ is 3-regular, i.e., all nodes are 3-valent,
the gauge-invariant spin-network basis $\{s_{\Gamma,j}\}$ is given
by contraction of the link representations $\{j_{i}\}$ with the unique
trivial intertwiners, i.e., Clebsch--Gordan symbols.

In 2+1 dimensions the spin-network basis diagonalizes the length
operators with a spectrum given in terms of the $SU(2)$ Casimir operator \cite{BenAchour:2013wx,laurentcarloetera}:
\begin{equation}\label{spede}
\widehat{l_{i}^{2}}|s\rangle=l_{i}^{2}|s\rangle=l^{2}(j_{i})|s\rangle=l_{\gamma}^{2}C_{j_{i}}|s\rangle=l_{\gamma}^{2}[j_{i}(j_{i}+1)+c]|s\rangle\,,
\end{equation}
where the scale $l_{\gamma} = l_{\rm Pl} \gamma$ is set by the Planck length and the Barbero--Immirzi parameter and $c$ is a constant dependent on the quantization map chosen for the Casimir operator \cite{laurentcarloetera,fluxmap}.
%In the following we use the freedom in choosing a constant $c$ setting it to $c=1/4$.
If $c>0$, the Clebsch--Gordan conditions
\begin{equation}
\left|j_{1}-j_{2}\right|\leqslant j_{3}\leqslant j_{1}+j_{2},
\end{equation}
implicit in the intertwiners on vertices of $\Gamma$ yield triangle inequalities on the primal complex,
\begin{equation}\label{trianineq}
l_{1}+l_{2} > l_{3}\,.
\end{equation}
This can be seen from the inequality 
\begin{equation}
\sqrt{j_{1}(j_{1}+1)+c}+\sqrt{j_{2}(j_{2}+1)+c}>\sqrt{\left(j_{1}+j_{2}\right)(j_{1}+j_{2}+1)+c}\geqslant\sqrt{j_{3}(j_{3}+1)+c}.
\end{equation}
Only for $c = 0$ there are degenerate spin configurations where the inequality on the lengths is not strict (e.g., $(j_1,j_2,j_3) = (0,1,1)$).
Even in this case one could obtain the triangle inequalities by  restricting to those states in $\mathcal{H}_{\rm kin}$ corresponding to non-degenerate configurations. 
We do not expect this to be a significant restriction in the calculation of the quantum spectral dimension. In the following we take $c = 1/4$, although calculations of the spectral dimension with different values indicate that $\ds$ is not so much sensitive to such choice. In fact, we have tested the $c=0$ case with very similar results.

Now, an obvious choice is to take an explicit form of the discrete Laplacian in terms of edge-length variables (see \cite[equation (63)]{COT1}). As an operator, it acts on a spin-network state $\{s_{\Gamma,j}\}$ as 
\begin{equation}
\widehat{\Delta}=\widehat{\Delta(l_i^{2})}=\Delta(\widehat{l_i^{2}})=\Delta(C_{j_i})=\Delta(j_i)\,.
\end{equation}
While the form of the primal volumes in the Laplacian, i.e., the triangle areas $A_{\sigma}(l_{i}^{2})$ and the edge lengths themselves, is straightforward, 
there is in principle some freedom in choosing the form of the dual edge lengths as functions of primal lengths.
We choose an expression related to a barycentric dual as argued for above and in \cite{COT1}.

%\

Let us now check whether the formal expression for the heat-trace
expectation value is well defined in this context, i.e., if the heat
trace is a self-adjoint operator. More precisely, the question is the following: Is the heat trace operator  $\widehat{P(\tau)}=\widehat{\Tr\,\rme^{\tau\Delta}}$
self-adjoint on the kinematical Hilbert space $\mathcal{H}_{\rm kin}$?
That is, for $s,s'\in\mathcal{H}_{\rm kin}$,
\begin{equation}\label{usef}
\langle s|\widehat{P(\tau)}s'\rangle\stackrel{?}{=}\langle\widehat{P(\tau)}s|s'\rangle\,.
\end{equation}
%Self-adjointness of $\widehat{P(\tau)}$ is a necessary precondition for any concept of a spectral dimension of quantum geometry.
With the above Laplacian, the left-hand side is
\begin{equation}
\langle s_{\Gamma,j}|\widehat{P(\tau)}s_{\Gamma',j'}\rangle=\langle s_{\Gamma,j}|\Tr\,\rme^{\tau\widehat{\Delta}}s_{\Gamma',j'}\rangle=\Tr\,\rme^{\tau\Delta(j)}\delta_{\Gamma,\Gamma'}\delta_{j,j'}\,,
\end{equation}
where $\delta_{\Gamma,\Gamma'}$ is the identity of graphs up to automorphisms (taking the usual inner product of LQG, states defined on distinct graphs are orthogonal). This is equal to the right-hand side of (\ref{usef}) if, and only if, the spectrum of $P(\tau)$ is real:
\begin{equation}
\langle\widehat{P(\tau)}s_{\Gamma,j}|s_{\Gamma',j'}\rangle=\langle\Tr\,\rme^{\tau\widehat{\Delta}}s_{\Gamma,j}|s_{\Gamma',j'}\rangle=[\Tr\,\rme^{\tau\Delta(j)}]^{*}\delta_{\Gamma,\Gamma'}\delta_{j,j'}\,.
\end{equation}
%This argument works for any classical observables in terms of which the Laplacian can be expressed and which have quantum observables diagonal in some orthonormal basis of $\mathcal{H}_{kin}$. E.g. for holonomy or flux bases if $\Delta$ can be expressed of holonomies of fluxes. 

In general, since the entries of $\Delta$ consist of lengths and areas, they are real on geometric states where triangle inequalities (\ref{trianineq}), or equivalently closure constraints (which are geometricity conditions) are satisfied. Then, $\Tr\,\rme^{\tau\Delta}$ is real as well.
Hence $\widehat{P(\tau)}$ is a good quantum observable on the kinematical states of 2+1 gravity with the operator ordering of $\hat{l}$ chosen \cite{laurentcarloetera,fluxmap} such that $c>0$, which we are considering here.
%or, alternatively, for $c=0$, if a restriction is imposed on those states in $\mathcal{H}_{\rm kin}$ corresponding to non-degenerate configurations. We do not expect this to be a significant restriction in the calculation of the quantum spectral dimension. In the following we take $c = 1/4$, although calculations of the spectral dimension with different values indicate that $\ds$ is not so much sensitive to such choice. In fact, we have tested the $c=0$ case with very similar results.

%Furthermore, because of the form of the heat trace as a trace over an operator exponential $\widehat{P(\tau)}=\Tr e^{\tau\widehat{\Delta}}$ which, explicitly, amounts to a sum over exponentials, the heat traceis positive when real. Thus, the heat trace expectation value on these states can indeed be given a probability interpretation as usual in the random walk picture.
%%% - One has further to check if the integral over $\tau$ converges such that the heat trace is indeed normalizable
%\ 

So far, we have given and discussed the definition of spectral dimension in the smooth, discrete and quantum cases, in particular for quantum states of geometry in 2+1 LQG, thereby setting the stage for its calculation. Since the quantum properties of such observable depend on their interplay with classical topological and discreteness effects, in the next two sections we will first investigate these effects separately before addressing the quantum case.

%%%%%%%%%%%%%%%%%%%%%%%%%%%%%%%%%%%%%%%%%%%%%%%%%%%%%%%%%%%%%%%%%%%%%%%%%%%%%%%%%%%%%%%%%%%%%%%%

\section[Simple examples for effects of topology and geometry: \texorpdfstring{$S^d$}{} and \texorpdfstring{$T^d$}{}]{Simple examples for effects of topology and geometry: sphere and torus}\label{sec:Topo}

In this section, we calculate the spectral dimension of spheres and tori and discuss the role of topology and geometry in these examples. In preparation for the actual LQG calculations (where these topologies are chosen), it is important to have these effects under control.

At large diffusion scales $\tau$, the dominant effects in the behaviour of the heat trace and the spectral dimension are due to the topology
of $M$. 
%When interested in the geometric information of quantum states encoded in the spectral dimension later it is important to know these effects to subtract them out. 
Qualitatively, a compact topology leads to a fall-off of $\ds(\tau)$
to zero. The scale at which this happens is related to the geometry
(i.e., the curvature radii). In the light of the heat trace interpretation in terms of random walks, this can be easily understood: the
return probability approaches one after diffusion times $\tau$ larger than the circumferences (closed geodesics). The resolution interpretation is also easy to spell out: as the resolution is lowered, one cannot distinguish different points but, due to the limited compact geometry of the set, the particle does not spread over an infinite manifold and it roughly appears localized in the same region.

To clarify these topological and geometric features, we discuss
the case of the circle $S^{1}$ and the generalization to the $d$-torus
$T^{d}$ and the $d$-sphere $S^{d}$. For $T^{d}$, we even find analytic solutions in closed form for the heat trace.

%\

\begin{figure}
\centering
\includegraphics[width=7.5cm]{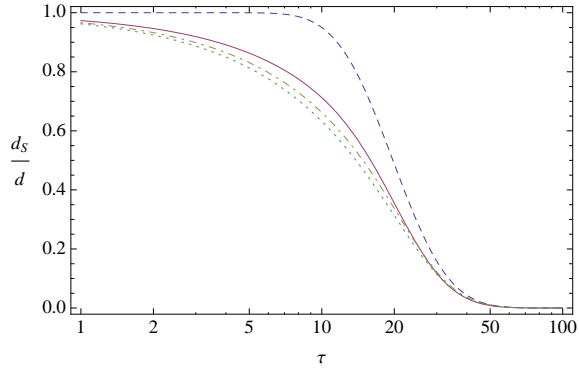}
\caption{Spectral dimension of $S^{d}$ rescaled to volume $V_{S^{d}}=1$,
for comparison divided by $d$, for $d=1,2,3,4$ ($S^1$ dashed line, $S^2$ solid, $S^3$ dot-dashed and $S^4$ dotted). The inverse square of the corresponding radii sets the scale of the topological effect. 
At the same time, it is a comparison with the case of $T^{d}$ with radii $R=1/(2\pi)$ as these tori are equivalent to the case of $S^{1}$.
\label{fig:dSsmooth}}
\end{figure}

On a circle of radius $R$, the spectrum of the Laplacian is $-(k/R)^{2}$, $k\in\Z$, yielding a heat trace
\begin{equation}
P_{S^{1}}(\tau)=\underset{k\in\Z}{\sum}\rme^{-\left(\frac{k}{R}\right)^{2}\tau}=\theta_{3}\left(0,\,\rme^{-\frac{1}{R^2}\tau}\right)=\theta_{3}\left(0\Big|\,\frac{1}{R^2}\frac{\rmi\tau}{\pi}\right)
\end{equation}
in terms of the third theta function\footnote{Here and in the following we use the notation for theta functions as defined in \cite{NIST}.}.
While $\ds(\tau)=d=1$ for $\tau<R{}^{-2}$, due to the periodicity
there is a fall-off to zero for $\tau>R{}^{-2}$ (figure \ref{fig:dSsmooth}).

The geometry defined by the constant curvature radius $R$ governs only the scale at which the topological effect takes place. In that sense, the fall-off of the spectral dimension  can be understood as due to the combination of geometry and topology.

\subsection{The torus \texorpdfstring{$T^{d}$}{}}

\begin{figure}
\centering
\includegraphics[width=7.5cm]{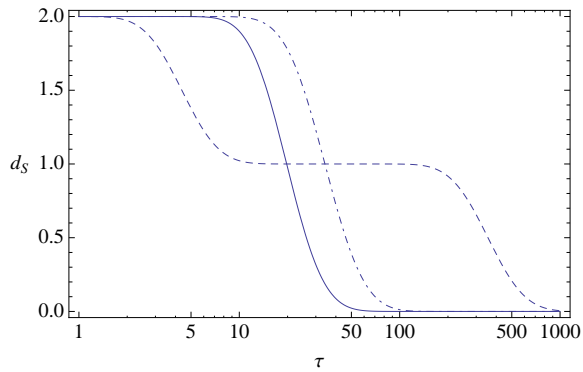}
\caption{Spectral dimension of $T^2$ with various geometries, $(R_1,R_2)=(1,1)$, $(R_1,R_2)=(1,\sqrt{3}/2)$, $(R_1,R_2)=(3,1/3)$ (solid, dash-dotted and dashed curve)}
\label{fig:dsT2}
\end{figure}

The torus $T^{d}=\left(S^{1}\right)^{\times d}$ with radii $R_{i}$
generalizes the case of the circle straightforwardly with spectrum
$-\sum_{i=1}^d k_i/R_i$, % $$-\overset{d}{\underset{}}\frac{k_{i}}{R_{i}}$,
 $\vec{k}\in\Z^{d}$, such that
\begin{equation}
P_{T^{d}}(\tau)=\underset{\vec{k}\in\Z^{d}}{\sum}\rme^{-\sum\left(\frac{k_{i}}{R_{i}}\right)^{2}\tau}=\theta\left(0\,\Bigg|\,\frac{\rmi\tau}{\pi}\left(\begin{array}{ccc}
R_{1}^{-2}\\
 & \ddots\\
 &  & R_{d}^{-2}
\end{array}\right)\right),
\end{equation}
where $\theta$ is the (multi-dimensional) Riemann $\theta$-function \cite{NIST}.
The spectral dimension of the case where all radii are equal turns
out to be just $d$ times the spectral dimension of $T^{1} = S^1$, since 
\begin{equation}
P_{T^{d}}(\tau)\propto\left[P_{S^{1}}(\tau)\right]^{d}.
\end{equation}
On the other hand, with more geometric parameters given by the constant curvature radii $R_{i}$, one can see that the geometry affects not only the scale at which the decay starts, but also accounts for various geometric regimes (figure \ref{fig:dsT2}). If the radii are ordered as $R_1\geqslant R_2\geqslant\dots\geqslant R_d$, the spectral dimension is constantly the topological dimension for $\tau<1/R_1^2$ and zero for $\tau>1/R_d^2$. In intermediate regimes, it can have plateaux at heights corresponding to intermediate integer dimensions, if the radii are of sufficiently different order. This can be easily understood: When $k$ radii are much smaller than a given scale, the $d$-torus effectively appears as a $(d-k)$-torus for the diffusion process at that scale. 

\subsection{The sphere \texorpdfstring{$S^{d}$}{}}

The spectrum of the Laplacian acting on functions on the sphere $S^{2}\cong SU(2)/U(1)$ of radius $R$ is proportional to the $SU(2)$ Casimir, $-C(j)/R^{2}=-j(j+1)/R^{2}$, for $j\in\N$ with $(2j+1)$-fold degeneracy, leading to a heat trace \cite{Benedetti:2009fo}
\begin{equation}
P_{S^{2}}(\tau)=\underset{j=0}{\overset{\infty}{\sum}}(2j+1)\,\rme^{-\frac{j(j+1)}{R^{2}}\tau}.
\end{equation}
More generally, for the $d$-sphere $S^{d}$, $d\geqslant 1$, the spectrum
is given by the eigenvalues $-j(j+d-1)/R^{2}$ with multiplicities such
that
\begin{equation}
P_{S^{d}}(\tau)=1+\underset{j=1}{\overset{\infty}{\sum}}\left[\binom{d+j}{d}-\binom{d+j-2}{d}\right]\,\rme^{-\frac{j(j+d-1)}{R^{2}}\tau}\,.
\end{equation}
The spectral dimension of $S^{d}$ has a slower fall-off in comparison with the torus $T^{d}$, with $d\geqslant 2$.
Thus, $d$-spheres are examples where the topological dimension is only obtained in the limit $\tau \ra 0$.
In contrast with the torus, spheres do not exhibit any multi-scale behaviour since they are governed by only one geometric parameter.

%\

In this section, we have discussed the generic effects of spherical and toroidal topology and of the geometric curvature parameters. All compact topologies have a spectral dimension going to zero at scales of the order of the curvature radii. Below this scale, $\ds$ agrees with the topological dimension in the case of the tori, while for spheres such an accordance holds only in the small-scale (or infinite resolution) limit $\tau\to 0$ (figure \ref{fig:dSsmooth}). Furthermore, in general a $d$-torus is given by $d$ geometric parameters setting the intermediate scales at which the torus has an effective lower-dimensional behaviour (figure \ref{fig:dsT2}).

%%%%%%%%%%%%%%%%%%%%%%%%%%%%%%%%%%%%%%%%%%%%%%%%%%%%%%%%%%%%%%%%%%%%%%%%%%%%%%%%%%%%%%%%%%%%%%%%

\section{Discreteness effects \label{sec:dsDiscrete}}

Now we consider classical effects of space discreteness in the spectral dimension. To this aim, we calculate the spectral dimension of various examples of hyper-cubulations and triangulations. This is preliminary to the quantum case in a twofold way. First, it is important for the distinction of effects of the underlying discrete structure from additional quantum effects. Second, from a practical perspective, one can then choose those complexes with smaller and more controllable discreteness features. They are more appropriate for analyzing quantum effects of states thereon. We find that regular torus triangulations are the best candidates for this purpose.

\subsection{Hypercubic lattice\label{sub:Hypercubic}}

As a simple, purely combinatorial case to start with, we consider finite and infinite hypercubic lattices.

On a toroidal lattice $(\Z_{n})^{d}$, the spectral dimension can be calculated analytically. For the cycle graph $C_{n}$, which corresponds to the $d=1$ case $\Z_{n}$, the eigenvalues of the Laplacian are known to be \cite{Chung:1997tk}
\begin{equation}\label{distor}
\lambda_{k}=1-\cos\left(\frac{2\pi k}{n}\right)=2\sin^{2}\left(\frac{\pi k}{n}\right),\qquad k=1,\dots,n\,,
\end{equation}
such that
\begin{equation}
P^{d=1}(\tau)=\rme^{-\tau}\sum_{k=1}^{n}\rme^{\tau\cos\left(\frac{2\pi k}{n}\right)}
\end{equation}
and 
\begin{equation}
\ds^{d=1}(\tau)=2\tau\left[1-\frac{\sum_k\cos\left(\frac{2\pi k}{n}\right)\,\rme^{\tau\cos\left(\frac{2\pi k}{n}\right)}}{\sum_k \rme^{2\tau\cos\left(\frac{2\pi k}{n}\right)}}\right].
\end{equation}
Using trigonometric relations, a given sum over exponentials of cosines can be further simplified to a sum over hyperbolic cosines; e.g., for $n=8$
\begin{equation}
d_{S}^{d=1}(\tau)=2\tau\left[1-\frac{\sinh\left(2\tau\right)+\sqrt{2}\sinh\left(\sqrt{2}\tau\right)}{1+\cosh\left(2\tau\right)+2\cosh\left(\sqrt{2}\tau\right)}\right].
\end{equation}
As in the case of smooth tori, for the general case of arbitrary dimension $d$ 
\begin{equation}
P^{d}(\tau)=\overset{d}{\underset{j=1}{\prod}}P^{d=1}(\tau)\,,
\end{equation}
yielding just a pre-factor of $d$ for the spectral dimension:
\begin{equation}
\ds^{d}(\tau)=d\, \ds^{d=1}(\tau).
\end{equation}

This can be compared with an infinite lattice $\Z^{d}$. For $d=1$,
because of translational symmetry one can consider only the heat kernel $\phi_{n}=K_{0n}$ for the sites $(0,n)$,
satisfying \cite{Ambjorn:1999in}
\begin{equation}
\frac{\rmd\phi_{n}}{\rmd\tau}=\frac{\phi_{n+1}+\phi_{n-1}-2\phi_{n}}{2},\label{lattice1d}
\end{equation}
with initial condition $\ensuremath{\phi_{n}(0)=\delta_{n0}}$. The solutions are \cite{Ambjorn:1999in}
\begin{equation}
\phi_{n}(\tau)=\rme^{-\tau}I_{n}(\tau)\,,
\end{equation}
where $I_{n}(\tau)$ is the hyperbolic Bessel function. One can use translational symmetry ($K_{mn}=K_{m+i,n+i}$) to get the return probability
\begin{equation}
P(\tau)=\mbox{Tr}K(\tau)\propto K_{00}(\tau)=\phi_{0}(\tau)=\rme^{-\tau}I_{0}(\tau)\sim\frac{1}{\tau^{1/2}}[1+\mathcal{O}(1/\tau)].
\end{equation}
Therefore, only for $\tau\gg1$ the topological dimension $\ds=1$ is obtained, while in general 
\begin{equation}
\ds(\tau)=2\tau\left[1-\frac{I_{1}(\tau)}{I_{0}(\tau)}\right],\label{eq:ds1d}
\end{equation}
with a maximum of $d_{\rm S,max}\approx 1.22$ at $\tau\approx1.70$. 

We can generalize the solution of \cite{Ambjorn:1999in} to $d$-dimensional lattices where the heat equation is
\begin{equation}
\frac{\rmd\phi_{n}}{\rmd\tau}=\underset{i=1}{\overset{d}{\sum}}\frac{\phi_{n+e_{i}}+\phi_{n-e_{i}}-2\phi_{n}}{2d}\,,
\end{equation}
for a generic lattice labeling $n$ and shifts $\pm e_{i}$ in the $i$-direction. A more useful labeling is a multi-index $n=n_{1}n_{2}...n_{d}$:
\begin{equation}
\frac{\rmd\phi_{n_{1}...n_{d}}}{\rmd\tau}=\underset{i=1}{\overset{d}{\sum}}\frac{\phi_{n_{1}...n_{i}+1...n_{d}}+\phi_{n_{1}...n_{i}-1...n_{d}}-2\phi_{n_{1}...n_{d}}}{2d}\,.
\end{equation}
The solution is 
\begin{equation}
\phi_{n_{1}...n_{d}}(\tau)=\rme^{-\tau}\underset{i=1}{\overset{d}{\prod}}I_{n_{i}}\left(\frac{\tau}{d}\right)\,;
\end{equation}
in particular,
\begin{equation}
K_{00}(\tau)=\phi_{00...0}(\tau)=\rme^{-\tau}\underset{i=1}{\overset{d}{\prod}}I_{0}\left(\frac{\tau}{d}\right)=\left[\rme^{-\frac{\tau}{d}}I_{0}\left(\frac{\tau}{d}\right)\right]^{d}.
\end{equation}
The heat trace scales as $d$ times the heat trace of the one-dimensional case. 

\begin{figure}
\centering
\includegraphics[width=7.5cm]{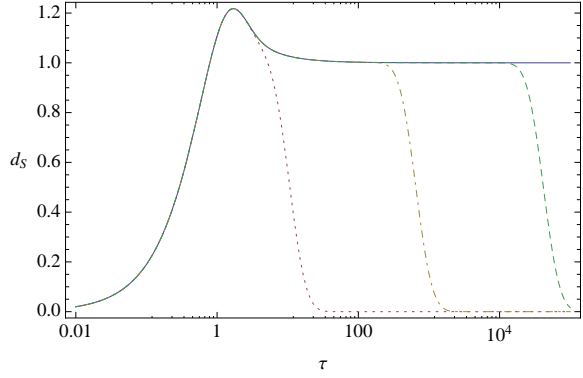}
\caption{Analytic $\ds$ for infinite $1d$ lattice, compared with the finite $\Z_{8}$ (dotted curve), $\Z_{64}$ (dot-dashed) and $\Z_{512}$ (dashed).}
\label{fig:dsZn}
\end{figure}

The small-$\tau$ behaviour of the finite lattice is exactly the same, as one can easily see numerically (figure \ref{fig:dsZn}).

Extending the discrete spectrum (\ref{distor}) to one parametrized by a continuous parameter $p\in[0,\pi[\In\R$ in the same interval,
\begin{equation}
2\sin^{2}\left(\frac{\pi k}{n}\right)\to2\sin^{2}p\,,
\end{equation}
the same result can also be shown by computing the heat trace analytically:
\begin{equation}
P(\tau)=\mbox{Tr}\,\rme^{-\tau2\sin^{2}p}\propto\int_{0}^{\pi}\rmd p\,\rme^{-\tau2\sin^{2}p}=\rme^{-\tau}I_{0}(\tau)\,.
\end{equation}
The straightforward extension to $d$ dimensions is given by
\begin{equation}
\Delta(p)%=\underset{i=1}{\overset{d}{\sum}}2\left(\frac{2}{a}\sin\frac{ap_{i}}{2}\right)^{2}\overset{a=2}{=}
=\underset{i=1}{\overset{d}{\sum}}2\sin^{2}p_{i}\,,
\end{equation}
so that
\begin{align}
P(\tau) & =\mbox{Tr}\,\rme^{-\tau\underset{i=1}{\overset{d}{\sum}}2\sin^{2}p_{i}}\propto\int_{0}^{\pi}\rmd^{d}p\,\rme^{-\underset{i=1}{\overset{d}{\sum}}\frac{\tau}{d}2\sin^{2}p_{i}}\\
 & =\underset{i=1}{\overset{d}{\prod}}\left(\int_{0}^{\pi}\rmd p_{i}\,\rme^{-\frac{\tau}{d}2\sin^{2}p_{i}}\right)=\left[\rme^{-\frac{\tau}{d}}I_{0}\left(\frac{\tau}{d}\right)\right]^{d}.
\end{align}
Incidentally, the use of a continuous parameter $p$ allows one to connect these results to the continuous case, $\Z^{d}\ra\R^{d}$. Introducing a lattice spacing $a$ (compare with lattice gauge theory,
e.g., \cite{Wiese:2009ub}) compatible with this spectrum,
\begin{equation}
\Delta(p)=2\left(\frac{2}{a}\sin\frac{ap}{2}\right)^{2}\overset{a=2}{=}2\sin^{2}p\,,
\end{equation}
where now $p\in[0,\frac{2\pi}{a}[$, the usual continuum limit is
\begin{equation}
2\left(\frac{2}{a}\sin\frac{ap}{2}\right)^{2}\underset{a\ra0}{\ra}2p^{2}.
\end{equation}
The formul\ae\ for the heat trace and spectral dimension are modified accordingly.

\subsection{Simplicial complexes
\label{sub:Triangulations}
}

The type of combinatorial structures that are most relevant for the quantum-gravity applications below are simplicial
complexes. In order to avoid having to deal with the difficult problem of constructing
 quantum-geometry states on infinite spaces, we restrict to simplicial manifolds
with compact topology.

\subsubsection*{Triangulations of $S^{2}$.}

\begin{figure}
\centering
\includegraphics[width=7.5cm]{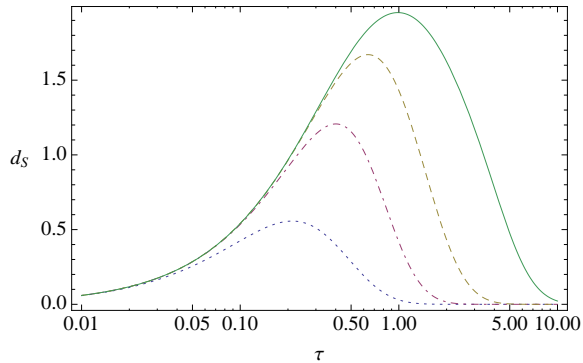}
\caption{$\ds$ of equilateral triangulations of $S^{2}$ in terms of the boundary of the (triangular) platonic solids, icosahedron, octahedron and tetrahedron (solid, dashed and dash-dotted curve) and of the dipole (dotted curve).  \label{fig:S2triang}}
\end{figure}

Obvious triangulations of the 2-sphere are the boundaries of the three
triangular platonic solids (tetrahedron, octahedron, icosahedron).
These are the only regular (i.e., equilateral) triangulations
of the smooth 2-sphere in terms of non-degenerate simplicial complexes.
The result for $\ds$ are shown in figure \ref{fig:S2triang}.

The larger the triangulation, the taller the height of the peak. In
particular, only the spectral dimension of the icosahedral
triangulation can be seen as providing a good approximation for the
topological dimension $d=2$, provided one defines it to coincide with the height of the peak (and assuming this would become a plateau for larger complexes, as seems to be the case in feasible calculations).

The extent to which too-small triangulations fail to capture the topological
dimension can be seen even more drastically in the case of the degenerate
complex, called dipole or (super)melon, triangulating a $d$-sphere in terms of just two simplices. Independently of the dimension $d$,
its heat trace is proportional to (see Appendix C in \cite{COT1})
\begin{equation}
P(\tau)=P(t=a\tau)=1+\rme^{-t},\label{eq:Pdipol}
\end{equation}
where the only geometric factor $a$ can be absorbed into the diffusion parameter. This yields a spectral dimension 
\begin{equation}
\ds(t)=\frac{2t}{1+\rme^t}\,.\label{eq:dSdipol}
\end{equation}
From its derivative 
\begin{equation}
\frac{\rmd}{\rmd t}\ds(t)=2\frac{1-\rme(t-1)\,\rme^{t-1}}{(1+\rme^t)^{2}}\,,
\end{equation}
it can be seen that the maximum is at $t^{\rm max}=W_{0}(1/\rme)+1\approx1.278$
(where $W_{0}$ is the upper branch of the real Lambert $W$-function) and has value
\begin{equation}
\ds^{\rm max}=\ds(t^{\rm max})\approx 0.56\,,
\end{equation}
 independent of the parameter $a$. Only the position of the maximum is rescaled by $a$.

\subsubsection*{Regular equilateral triangulations of $T^{d}$.}

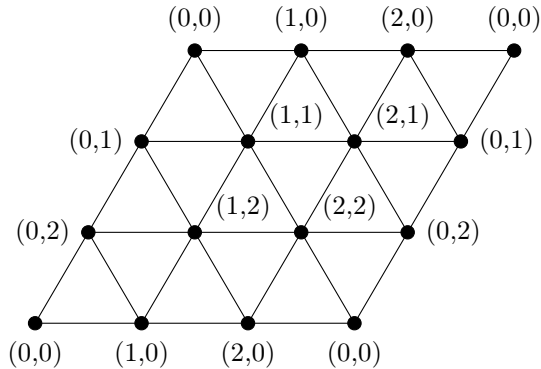
\begin{figure}
\centering
\begin{tikzpicture}[place/.style={circle,draw=black,fill=black, inner sep=0pt, minimum size=5},scale=0.7 ]
\draw (0,0) -- (6,0);
\draw (1,1.722) -- (7,1.722); 
\draw (2,3.444) -- (8,3.444); 
\draw (3,5.166) -- (9,5.166); 
\draw (0,0) -- (3,5.166); 
\draw (2,0) -- (5,5.166);
\draw (4,0) -- (7,5.166);
\draw (6,0) -- (9,5.166); 
\draw (2,0) -- (1,1.722); 
\draw (4,0) -- (2,3.444);
\draw (6,0) -- (3,5.166); 
\draw (7,1.722) -- (5,5.166);
\draw (8,3.444) -- (7,5.166);
\node at (0,0) [place, label=below:{\footnotesize(0,0)}]{}; 
\node at (2,0) [place, label=below:{\footnotesize(1,0)}]{}; 
\node at (4,0) [place, label=below:{\footnotesize(2,0)}]{}; 
\node at (6,0) [place, label=below:{\footnotesize(0,0)}]{}; 
\node at (1,1.722) [place, label=left:{\footnotesize(0,2)}]{};  
\node at (3,1.722) [place, label={[label distance=1pt]5:\footnotesize(1,2)}]{};  
\node at (5,1.722) [place, label={[label distance=1pt]5:\footnotesize(2,2)}]{};  
\node at (7,1.722) [place, label=right:{\footnotesize(0,2)}]{}; 
\node at (2,3.444) [place, label=left:{\footnotesize(0,1)}]{}; 
\node at (4,3.444) [place, label={[label distance=1pt]5:\footnotesize(1,1)}]{}; 
\node at (6,3.444) [place, label={[label distance=1pt]5:\footnotesize(2,1)}]{}; 
\node at (8,3.444) [place, label=right:{\footnotesize(0,1)}]{}; 
\node at (3,5.166) [place, label=above:{\footnotesize(0,0)}]{}; 
\node at (5,5.166) [place, label=above:{\footnotesize(1,0)}]{}; 
\node at (7,5.166) [place, label=above:{\footnotesize(2,0)}]{}; 
\node at (9,5.166) [place, label=above:{\footnotesize(0,0)}]{};
\end{tikzpicture}
\caption{Smallest regular (non-degenerate) simplicial complex of $T^{2}$ topology given by $p^2 = 3^2$ vertices
%, labeled by $p$-adic numbers 0 to 22, as natural and practical on $\Z_p^d$.
\label{fig:T2triang}}
\end{figure}

While there are no further, larger equilateral triangulations of $S^{2}$, for
the $d$-torus $T^{d}$ there are regular equilateral triangulations
of arbitrary combinatorial size (i.e., number of vertices) $N_{0}=p^{d}$. These are
obtained from hypercubic lattices via a so called standard triangulation of each hypercube
\cite{Brehm:2011kp} and are non-degenerate
for $p\geqslant 3$. In two dimensions, these are triangulations of the flat
torus with radii ratio $R_{1}/R_{2}=\sin(\pi/3)=\sqrt{3}/2$ (figure \ref{fig:T2triang}).

\begin{figure}
\centering
\includegraphics[width=7.5cm]{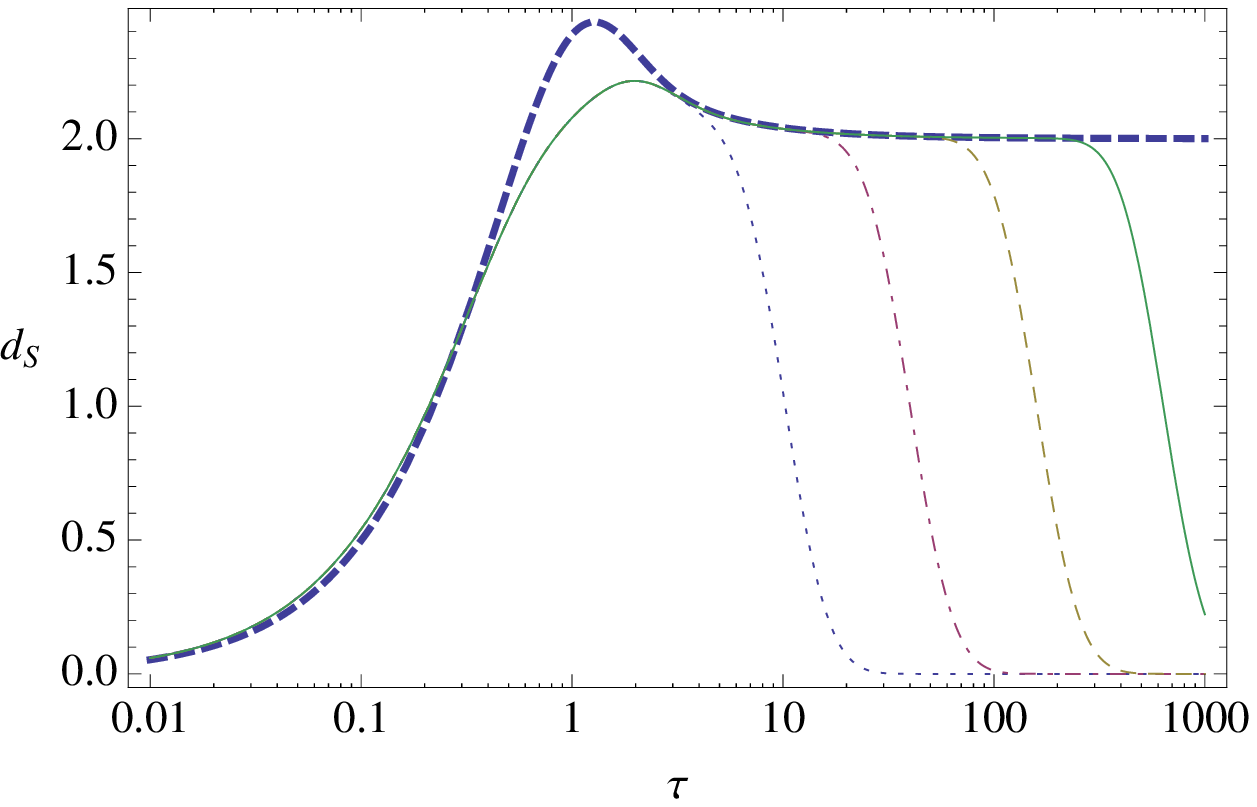}
\includegraphics[width=7.5cm]{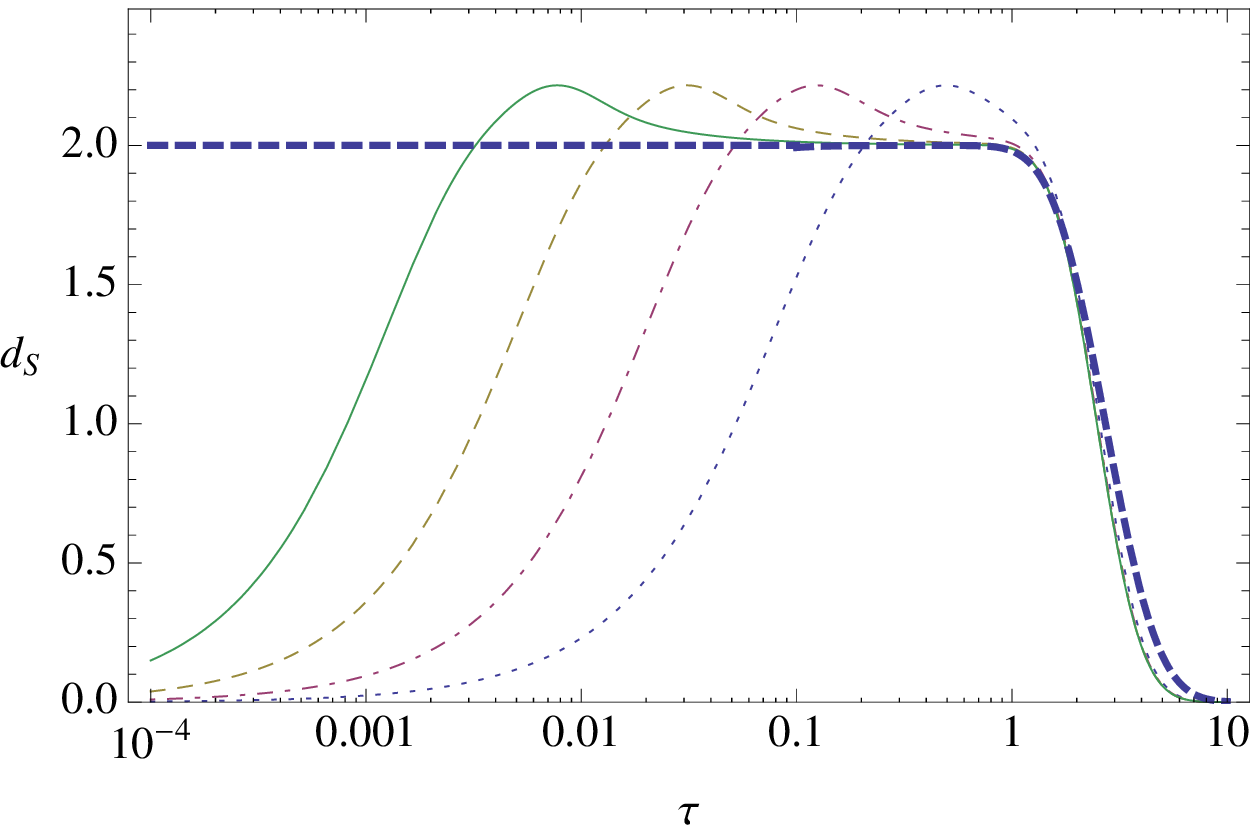}
\caption{Right plot: (a) unrescaled equilateral $T^{2}$ triangulation with $N_{0} = p^2 = (3\times 2^{k})^{2}$ vertices, $k=0,1,2,3$ from left to right (dotted, dot-dashed, dashed thin, solid curve)  indicating a convergence to the topological dimension for large $\tau$ in the $p\ra\infty$ limit, compared with the infinite quadrangular lattice (dashed thick).
Left plot: (b) rescaled triangulations ($k=0,1,2,3$ from right to left) indicating a convergence to the smooth $T^{2}$ (dashed thick) for $p\ra\infty$.
\label{fig:dsT2triang}}
\end{figure}

When comparing triangulations of different combinatorial size $N_{0}$,
one can either (a) fix the edge lengths to some scale $l_{*}$ such that the geometric size of the triangulations is growing with the combinatorial size, or one can (b) rescale them to $l^{(p)}=l_{*}/p$ according to the combinatorial size to keep the overall geometric size fixed. Thus, in the limit $p\ra\infty$ the former case gives a triangulated plane $\R^{2}$, while the latter approximates the smooth % triangulated
flat $T^{2}$ geometry.
Indeed, the calculations
in figure \ref{fig:dsT2triang} (for various finite $p$) indicate that
the spectral dimension of these triangulations capture both limits.

Moreover, these calculations show that the only difference between
a quadrangular ($2d$ hypercubic) and a triangular lattice of the plane consists in a
qualitatively different discretization effect given by the peak around
$\tau=1$.

\begin{figure}
\centering
\includegraphics[width=7.5cm]{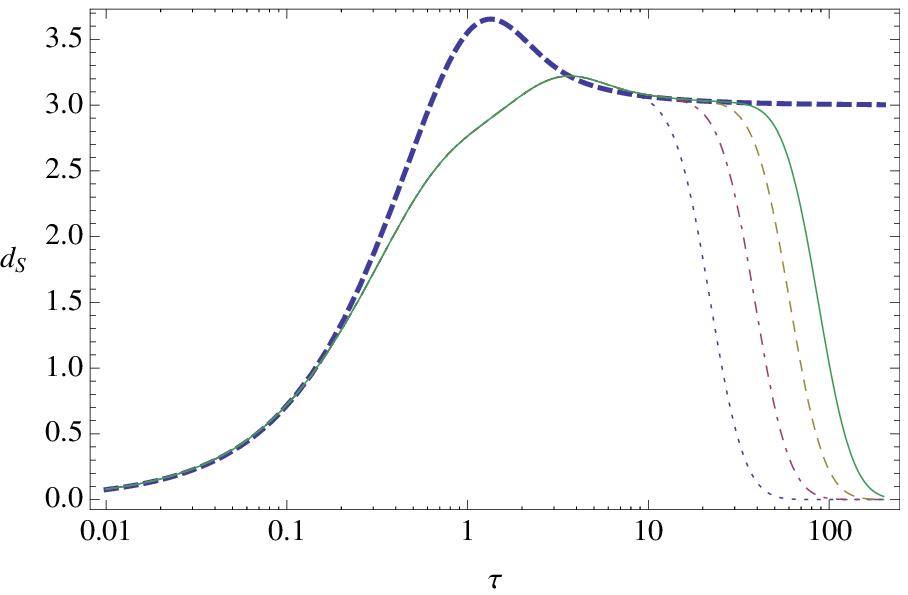}
\includegraphics[width=7.5cm]{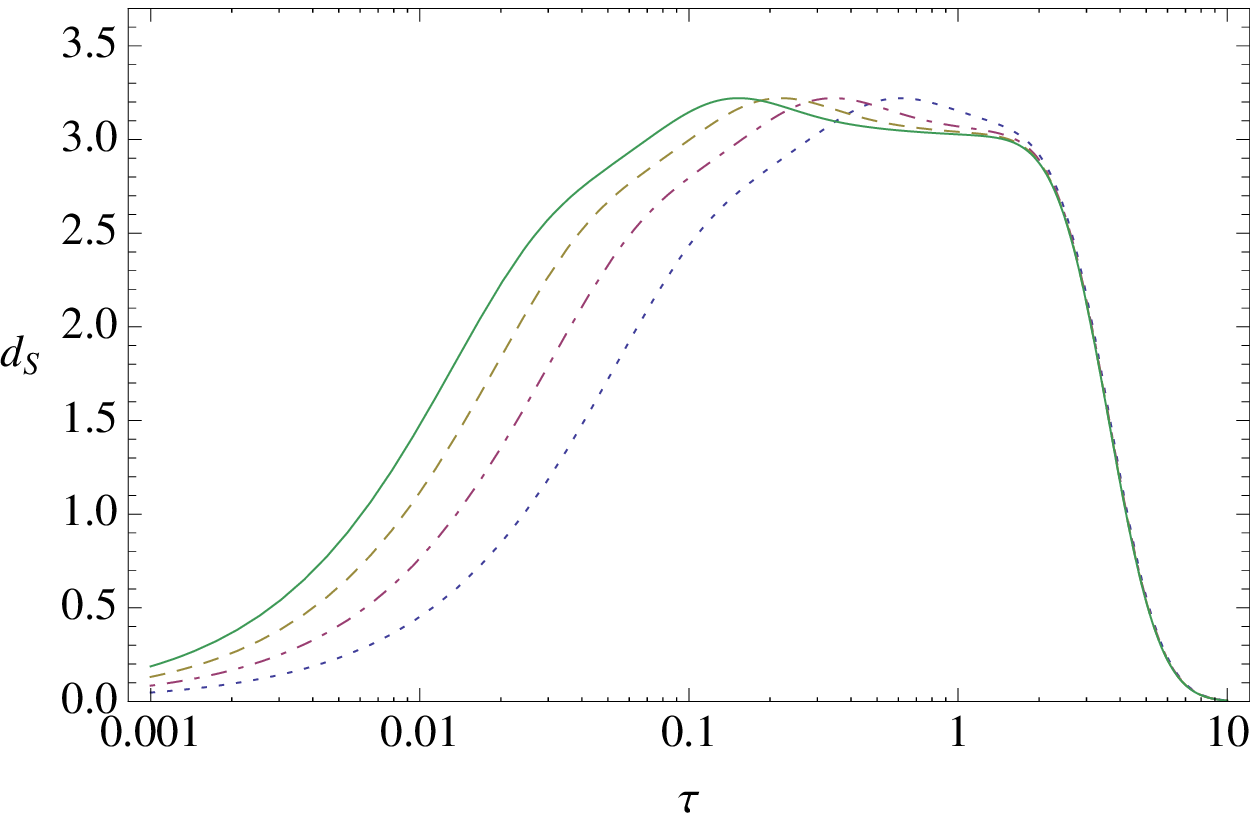}
\caption{Left: (a) equilateral $T^{3}$ triangulation in terms of $N_{0}=p^{3}$ vertices,
$p=6,8,10,12$ from left to right (dotted, dot-dashed, dashed thin, solid curve), compared with the cubic lattice (dashed thick). Right: (b) rescaled triangulations.\label{fig:dsT3triang}}
\end{figure}

The same analysis can be repeated in higher dimensions. In terms of
the standard triangulation of the cube, there is an equilateral triangulation
of the 3-torus with radii ratios $R_{2}/R_{1}=\sqrt{3}/2$ and $R_{3}/R_{1}\approx0.752$.
%(see Appendix \ref{sec:Constructing-complexes-and}) . 
The calculations in figure \ref{fig:dsT3triang} again indicate the correct behaviour in the limit
of infinite triangulations. The discretization effect here is slightly more marked, in that below the relatively small peak there is a small regime of relatively weak slope before the usual
steeper fall-off at small $\tau$ sets in. 

\subsubsection*{Subdivisions of triangulations of $T^{d}$.}

An alternative strategy for obtaining combinatorially larger simplicial manifolds of a given topology is to subdivide them. A natural way to do so is the Pachner 1-($d$+1) move where one $d$-simplex is subdivided into $d+1$ simplices by inserting a vertex in the middle of the original simplex and connecting it to all its vertices. 
%On the dual complex this amounts to resolving a vertex into a $d$-simplex.
%, which for $d=2$ looks similar to the combinatorial part of the proposed action of the LQG Hamiltonian \#\# 

Again, concerning the geometric realization of such a combinatorial structure, one can either (a) consider it as an equilateral triangulation (although this is not a triangulation of a torus with flat geometry anymore) or (b) rescale the edge lengths such that it is a triangulation of the flat torus.
One can, for example, consider the vertex $v$  inserted by the Pachner 1-($d$+1) move as the barycenter of a $d$-simplex with vertex set $(ijk\dots)$. The new edges $(iv)$ then need to have squared lengths
\begin{equation}
l_{iv}^{2}=\frac{1}{\left(d+1\right)^{2}}\left(d\underset{j}{\sum}l_{ij}^{2}-\underset{(jk)}{\sum}l_{jk}^{2}\right)
\end{equation}
to preserve the flat geometry approximated by the triangulation (see \cite{COT1}, Appendix B.2; sums are running over vertices $j$ and edges $(jk)$ of the simplex not containing the vertex $i$).

\begin{figure}
\centering
\includegraphics[width=7.5cm]{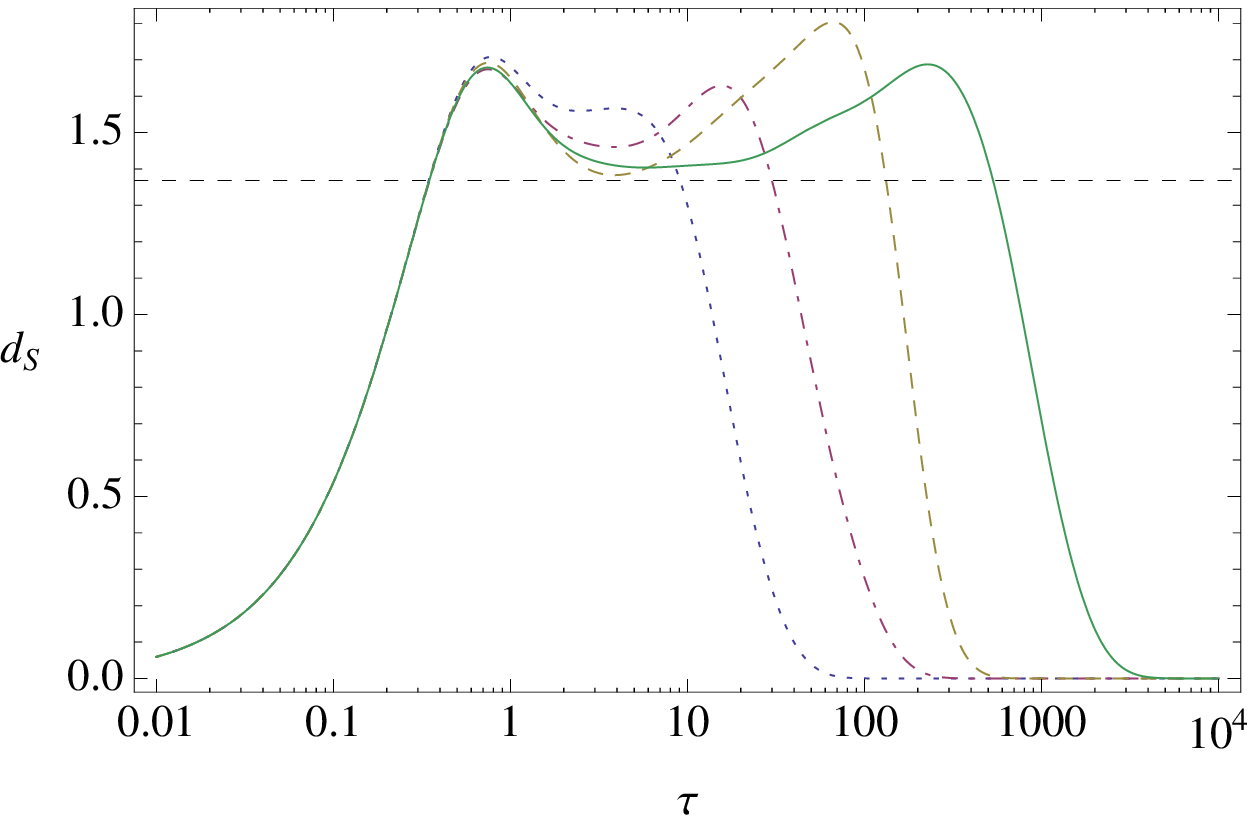}
\includegraphics[width=7.5cm]{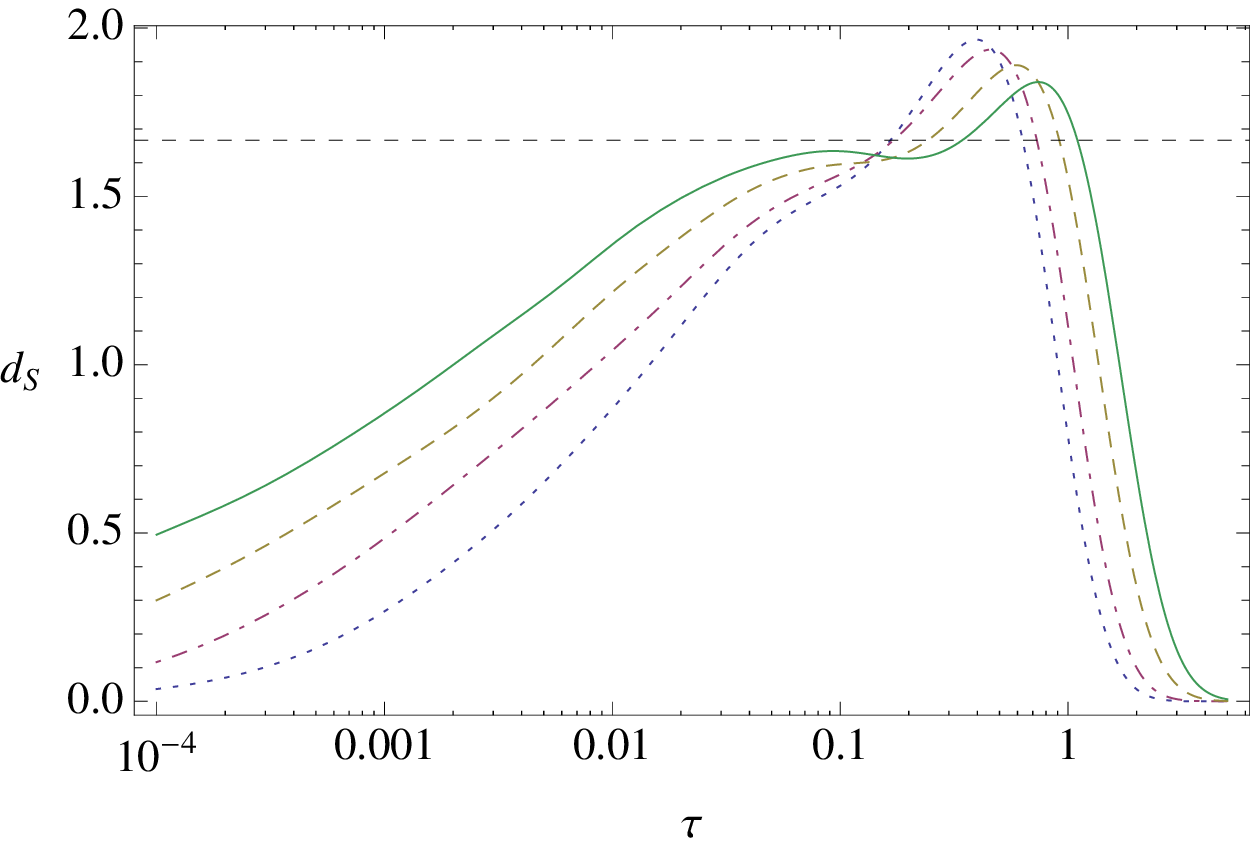}
\caption{Spectral dimension of the simplicial complex obtained from the %$N_{0}=9$ torus
smallest regular $T^{2}$-triangulation by $6\times 3^{k}$ single random subdivisions, $k=1,2,3,4$ (dotted, dot-dashed, dashed and solid curve)
% values chosen for comparison with the case of global subdivisions, figure \ref{fig:dsT2subdiv}),
taken (a) as equilateral (left, with dashed line at $d_S=1.37$) or (b) rescaled triangulations of the flat torus (right, with dashed line at $d_S=3/5$). \label{fig:dsT2subran}}
\end{figure}

We compare these two possibilities in two cases. (i) First, we subdivide the above $T^{2}$ triangulation ($N_{0}=9)$ applying the 1-($d$+1) move simultaneously on all triangles; (ii) second, we apply it randomly.
\begin{enumerate}
\item[(i)] In the first case of global subdivisions (figure \ref{fig:dsT2subdiv}), considered as equilateral triangulations, there is a peak slightly lower than $d=2$ at the diffusion time scale of the size of the triangulation. 
But at smaller intermediate scales we find small oscillations around a value of about $\ds\approx1.37$, obtained after integrating over a period. 
It is not surprising that there is a deviation from the topological dimension, since these equilateral triangulations have geometric realizations only in terms of a torus curved at various scales in a specific manner.

What might be more surprising is that also the complex with barycentrically rescaled edges triangulating the flat torus has a spectral-dimension plot quite different from the equilateral triangulations and cubulations considered above. 
There is a more complicated oscillatory behaviour but now at a value around $5/3$. 
Furthermore, the fall-off at small $\tau$ is much less steep. 

The rescaled Pachner subdivisions are thus an example of a triangulation which substantially deviates with respect to the spectral dimension in the corresponding continuum geometry, due to the particular combinatorial structure.
%Since these are still triangulations of the flat torus these effects in the spectral dimension are obviously due to the kind of triangulation used, and more precisely to both the combinatorial structure as well as the geometric data, i.e., edge lengths. 
In particular, in the finite case there is no plateau at the value of the topological dimension, nor is it expected that the topological dimension be recovered in the large-size limit.
An important lesson one can draw is that the spectral dimension of a triangulation of a given geometry does depend on the combinatorics of the chosen triangulation.
% with respect to the spectral dimension, to recover the dimension plot of a smooth geometry from a discrete one it is not enough to take the infinite size of the triangulation.
\begin{figure}
\centering
\includegraphics[width=7.5cm]{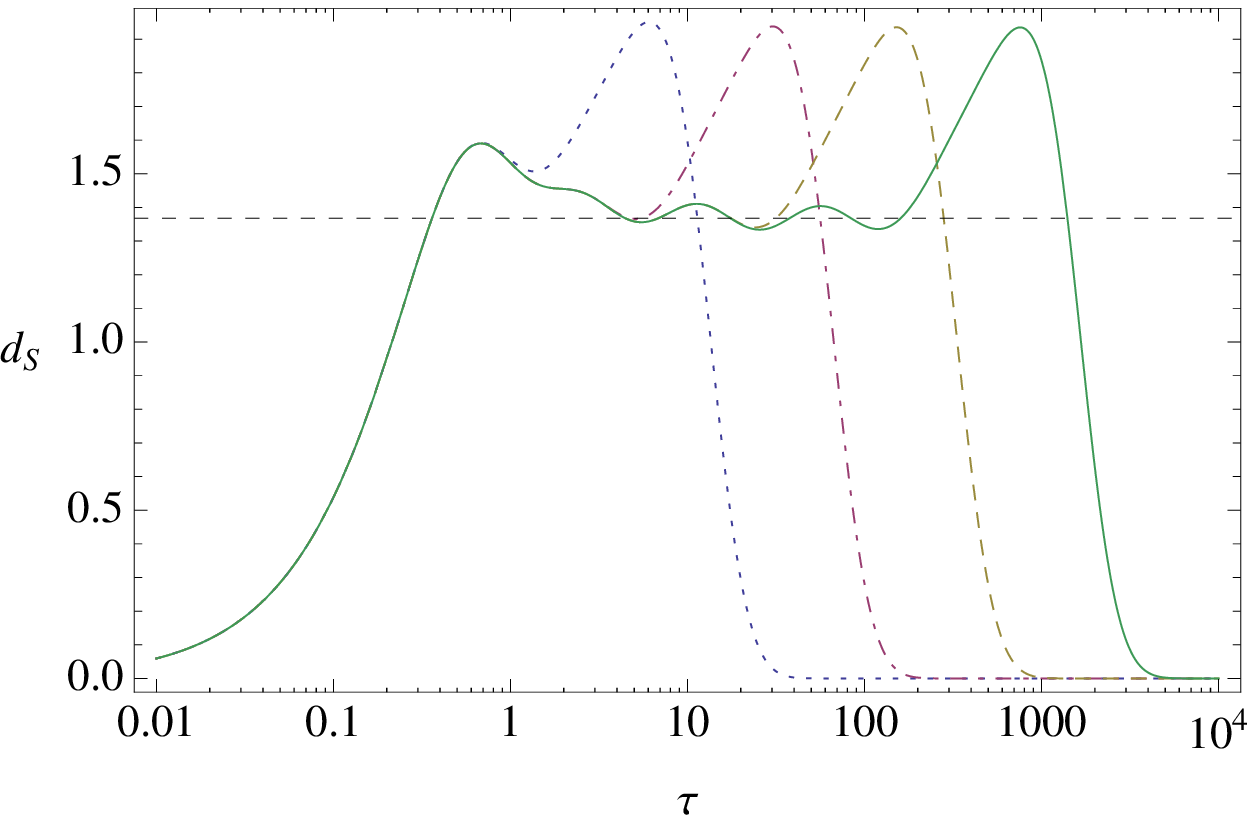}
\includegraphics[width=7.5cm]{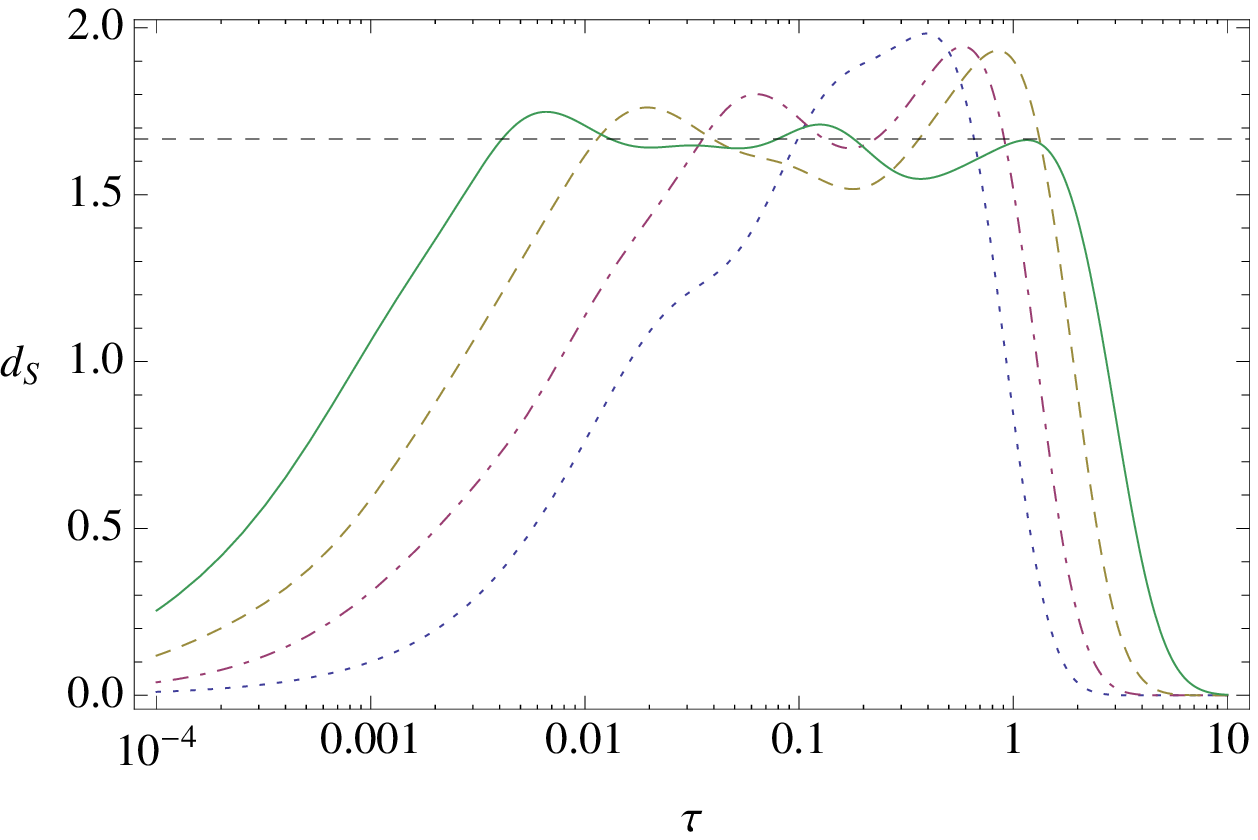}
\caption{Spectral dimension of $k$ global subdivisions of the smallest regular $T^{2}$-triangulation, $k=1,2,3,4$ (dotted, dot-dashed, dashed and solid curve).
Left: (a) purely combinatorial complex, with a dashed line at $d_S=1.37$. 
%($k$ increasing from left to right)
Right: (b) complex with rescaled edge lengths triangulating the flat torus, with a dashed line at $d_S=3/5$.}
\label{fig:dsT2subdiv}
\end{figure}

\item[(ii)] In the second case, we construct subdivisions by choosing (with uniform random distribution)
one triangle to subdivide, performing the subdivision, and then re-iterating the process. 
The results (figure \ref{fig:dsT2subran}) hint at some kind of averaging effect. 
In the equilateral case, the oscillations are washed away and the height of the peak at the volume-size scale seems to be dependent on the particular element chosen in the random ensemble. The closer is the element to a global subdivision, the more pronounced the peak.

In the rescaled case, the regime around $3/5$ is much smaller than in the global subdivided case and the fall-off is even less steep. This might be explained by the fact that the random subdivision effectively averages over both the regime corresponding to a plateau and the regime of low-$\tau$ fall-off.
\end{enumerate}

\begin{figure}
\centering
\includegraphics[width=7.5cm]{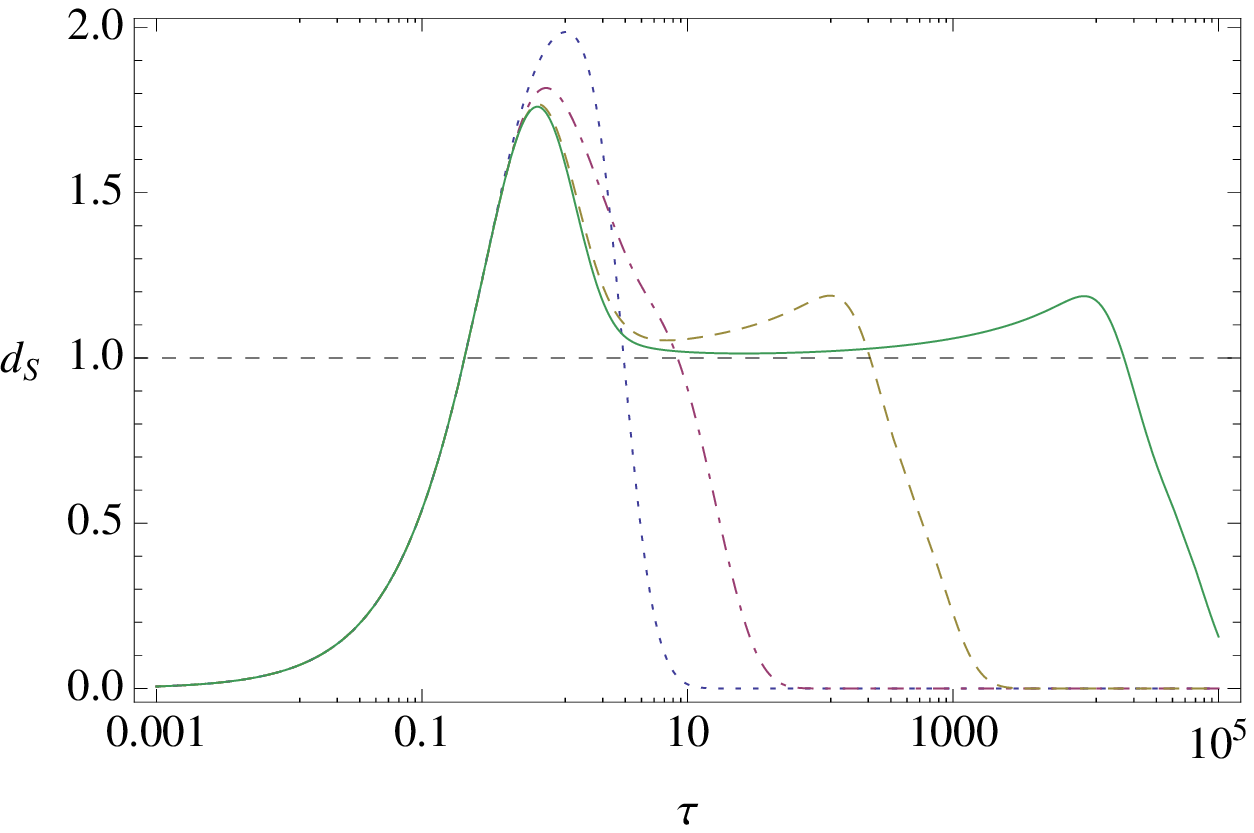}
\includegraphics[width=7.5cm]{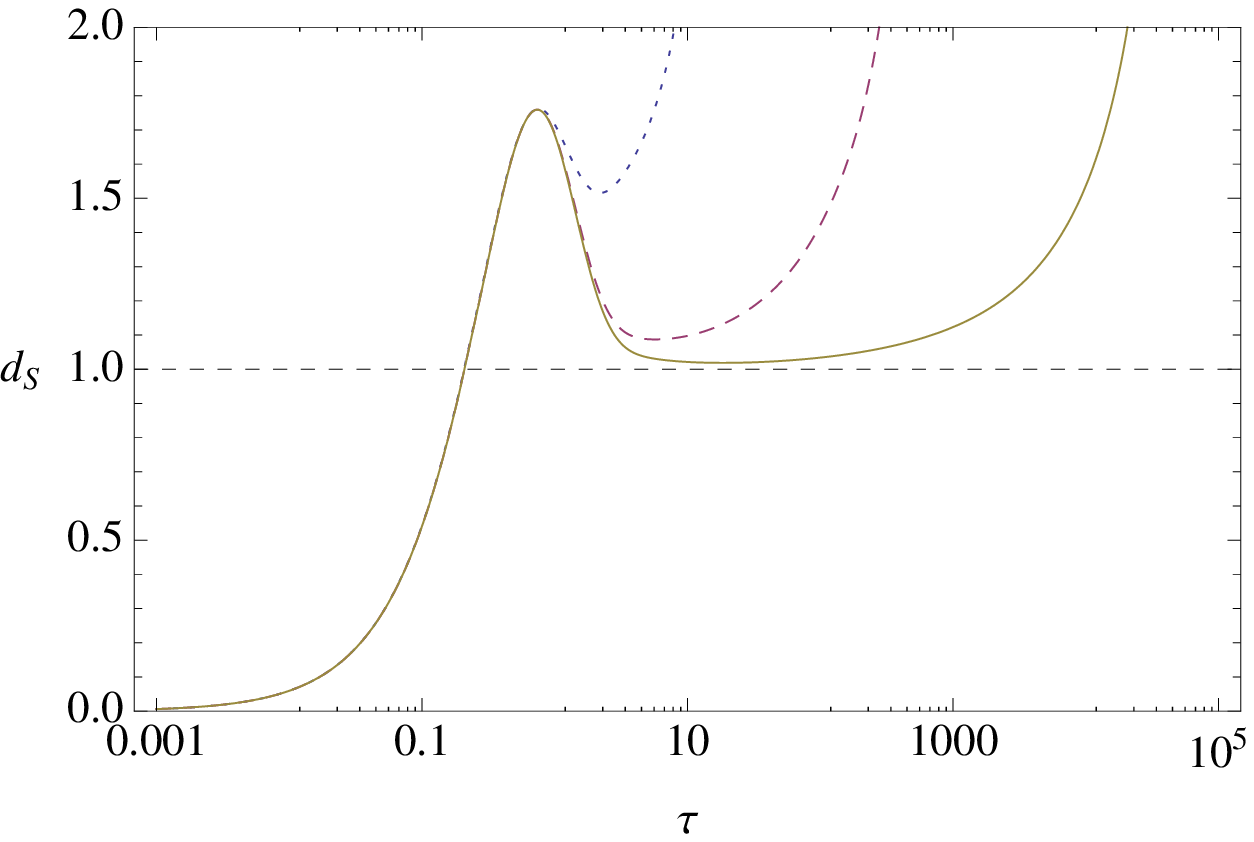}
\caption{Spectral dimension of the simplicial complex obtained by $10^{k}$ single subdivisions, $k=0,1,2,3$, around the same vertex of the $N_{0}=9$ torus triangulation (left) and of a single triangle (right), both unrescaled. In both cases, $\ds$ is independent of the global structure in the large complex size limit. (In the right plot, the spectral dimension goes to infinity for large $\tau$ since there is no zero eigenvalue in the spectrum of the Laplacian due to the boundary.)
\label{fig:dsT2frac}}
\end{figure}

%\
Finally, we have checked a very peculiar element in the random ensemble,
namely the repeated subdivision around the same vertex of the triangulation.
This is interesting because it shows that the spectral dimension
is very much dependent on the combinatorics, yielding a spectral dimension which could be conjectured to run to $\ds\ra1$ in the large-size limit, as suggested by the calculations at finite sizes (figure \ref{fig:dsT2frac}). Since this property of the example is actually independent of the global structure of the complex, one can have the same result starting with only one triangle.

\

In this section, we have investigated discreteness effects in the spectral dimension for various equilateral lattices. In general, there are three characteristic features. (A) A zero spectral dimension at scales below the lattice spacing, $\ds\simeq 0$, coming from the fact that the test particle feels the discrete spacing of the lattice: for too-short or infinitesimal times $\delta\tau$, the probe does not have the chance to diffuse from a given node to another. (B) A peak larger than the topological dimension at the lattice scale. (C) At larger scales, there is agreement with the smooth spectral dimension in the case of regular toroidal hypercubulations (figure \ref{fig:dsZn}) and triangulations (figure \ref{fig:dsT2triang}). In particular, there is a plateau with height close to the topological dimension, the more extended the larger the complex. This cannot be found for spherical equilateral triangulations, which only exist for certain sizes too small to establish a plateau (figure \ref{fig:S2triang}).  
%The regular lattices can be understood as a way to approximation either of an infinite lattice for fixed lattice spacing or of the smooth case for when the lattice spacing is rescaled. This very much favors such triangulations for the definition of quantum states semiclassical to smooth geometries, to be considered in the next section.
Triangulations obtained by subdivisions via Pachner 1-($d$+1)-moves do not reproduce the topological dimension neither in the unrescaled nor rescaled case (figures \ref{fig:dsT2subdiv} and \ref{fig:dsT2subran}). Thus, already at the classical level the spectral dimension is very sensitive to the precise structure of the discrete manifolds combinatorics.

%%%%%%%%%%%%%%%%%%%%%%%%%%%%%%%%%%%%%%%%%%%%%%%%%%%%%%%%%%%%%%%%%%%%%%%%%%%%%%%%%%%%%%%%%%%%%%%%

\section{Spectral dimension of states of quantum geometry (in \texorpdfstring{$d=2$}{})\label{sec:dsKinematics}}

After these preliminary considerations, we can now calculate the spectral dimension of various states of 2+1 Euclidean quantum geometry. First, we want to consider coherent states, that is states which are peaked both on an intrinsic, two-dimensional spatial geometry as well as on its conjugate extrinsic geometry. The aim is to identify the quantum corrections to the classical spectral dimension and find the dependence of $\ds$ on the parameters of the coherent states.
%  states randomized in the intrinsic geometry.
In these cases, our results show only qualitative and small deviations from the classical case. In an attempt to uncover stronger quantum effects, we finally investigate superpositions of coherent states. The resulting spectral dimension turns out to be an average of the superposed cases, thereby showing a more distinct quantum behaviour.

\subsection{Coherent spin-network states}

In the LQG literature, the starting point to get coherent LQG states is usually group coherent states in the holonomy basis. Coherent states peaked at phase space points $(g,x=0)\in T^*G\cong G\times\mathfrak{g}$
can be obtained from the heat kernel $K^{\sigma}$ on the group $G=SU(2)$ with Peter--Weyl expansion:
\begin{equation}
\psi_{(g,0)}^{\sigma}(h)=K^{\sigma}(hg^{-1})=\sum_{j}d_{j}\,\rme^{-\frac{C_{j}}{2 \sigma^2}}\chi^{j}(hg^{-1}).
\end{equation}
Here $d_j$ refers to the dimension and $\chi^j$ to the character of the $G$ representation labeled by $j$, $\sigma$ is a positive spread, $\sigma \in \R^+$. There are two known constructions for a generalization including a non-trivial $\mathfrak{g}$-dependence: Either by analytic continuation to $\rme^{\rmi x}g\in G^{\C}$ such that \cite{Hall:1994gg,Sahlmann:2001bw,Bahr:2009bc}
\begin{equation}
\psi_{(g,x)}^{\sigma}(h)=K^{\sigma}(hg^{-1}\rme^{-\rmi x})
\end{equation}
or using the flux representation \cite{Oriti:2012kx} where the heat kernel appears to be a ($\kappa$-non-commutative) Gaussian 
\begin{equation}
\tilde{\psi}_{(g,x)}^{\sigma}(y)=\mathcal{F}[\psi_{g}^{\sigma}](x-y)\propto \rme_{\star}^{-\frac{1}{2\kappa^{2}}\frac{(x-y)^{2}}{2\sigma^2}}\star \rme^{\frac{\rmi}{\kappa}|P(g)|(x-y)}.
\end{equation}
Here, $\kappa=l_{\rm Pl}^{-1}$ and plane waves $e_g(x):=\rme^{\frac{\rmi}{\kappa}|P(g)| x}$ use group coordinates $\vec{P}(g) = \sin[\theta(g)] \vec{n}$ in the usual coordinates in which the group element is parametrized as $g = \rme^{\rmi\theta \vec{n} \cdot \vec{\sigma}}$, with $\vec{\sigma}$ the Pauli matrices. The notation $\rme_\star$ indicates the non-commutative exponential defined as a power series expansion of $\star$-monomials \cite{fluxmap}. Transforming back
to group space, this results in just an additional plane wave factor,
\begin{equation}
\psi_{(g,x)}^{\sigma}(h)=\psi_{(g,0)}^{\sigma}(h)\,e_{h}(-x)\,.
\end{equation}

Since we are considering the Laplacian operator diagonal in the spin
representation, we need the spin expansion of these coherent
states. In both cases, there is a limit (for large enough spins) in
which these can be described \cite{Hall:1994gg,Sahlmann:2001bw,Oriti:2012kx,BMP} as Gaussian-type states peaked at spin
representation labels of intrinsic geometry $\{J_{l}\}$ and angles of extrinsic
geometry $\{K_{l}\}$:
\begin{equation}
|\psi_{\Gamma}^{\{J_{l},K_{l}\}}\rangle=\frac{1}{N}\sum_{\{j_{l}\}}\psi_{\Gamma}^{\{J_{l},K_{l}\}}(j_{l})|\Gamma,j_{l}\rangle\,,
\end{equation}
with spin-network basis coefficients 
\begin{equation}
\psi_{\Gamma}^{\{J_{l},K_{l}\}}(j_{l}) \propto \prod_{l\in\Gamma}\rme^{-\frac{(J_{l}-j_{l})^{2}}{2\sigma^{2}}+\rmi K_{l}j_{l}}.
\end{equation}
In fact, following \cite{BMP,Oriti:2012kx}, in the large-$x$ approximation, one finds that the $J_l$ can be identified (up to a factor dependent on $\sigma$) with the modulus $x$ of the flux,
%% $d_{J_l} = 2\sigma^2 x$ ??
and the $K_l$ are angles in the representation of the group elements $g_l$ (in the plane orthogonal to the flux $x$).
%% using above notation $K_l = \theta n_3$
For the details we refer to \cite{BMP} and \cite{Oriti:2012kx} respectively, since in the following only the intrinsic curvature as captured by the $J_l$ will be relevant. Their dependence on $\sigma$ does not play a role for fixed $\sigma$ or small variations of it, with respect to (assumed large) $x$.

The heat-trace expectation value is then
\begin{align}
\left\langle \widehat{P(\tau)}\right\rangle _{\Gamma}^{\{J_{l},K_{l}\}}
%% = \left\langle \widehat{\tilde{P}(\tau)}\right\rangle _{\psi_{\Gamma}^{\{J_{l},K_{l}\}}} 
& = \frac{1}{N^2} \underset{\{j_{l}\}}{\sum}\left|\psi_{\Gamma}^{\{J_{l},K_{l}\}}(j_{l})\right|^{2}\langle \Gamma,j_{l}|\mbox{Tr}\,\rme^{\tau\widehat{\Delta}}|\Gamma,j_{l}\rangle \\
& \propto \underset{\{j_{l}\}}{\sum}\left|\prod_{l\in\Gamma}\rme^{-\frac{(J_{l}-j_{l})^{2}}{2\sigma^{2}}+\rmi K_{l}j_{l}}\right|^{2}\mbox{Tr}\,\rme^{\tau\langle \Gamma,j_{l}|\widehat{\Delta}|\Gamma,j_{l}\rangle }\\
& =\underset{\{j_{l}\}}{\sum}\left[\prod_{l\in\Gamma}\rme^{-\frac{(J_{l}-j_{l})^{2}}{\sigma^{2}}}\right]\mbox{Tr}\,\rme^{\tau\Delta_{\Gamma}(j_{l})},\label{eq:htSuperposed}
\end{align}
from which the spectral dimension is derived according to equation (\ref{eq:qds}).
As the spatial Laplacian does not depend on the extrinsic curvature, it is natural that also the phase of the coherent state drops out of the expectation value.

%The heat trace and spectral dimension on coherent spin network states,thus, are effectively Gaussian averages peaked on the intrinsic geometry on $\Gamma$ given by the length spectra in terms of representations $\{J_{l}\}$.
%For coherent spin network states one would therefore expect the spectral dimension to be approximately equal to the one of the classical triangulation with geometry according to (the spectra) of $\{J_{l}\}$.
%\
%Thus, the relevant information in $d_{S}$ of a coherent state should consists in its difference to $d_{S}$ of the underlying simplicial complex. This difference can be seen as the quantum corrections of the semiclassical state.

The reason for considering coherent states here is twofold. First, as they are semi-classical states peaking at classical geometries, it is interesting to check if their properties are reflected in the spectral dimension. Second, if the quantum spectral dimension was comparable with the spectral dimension of the classical geometries peaked at, the difference should be understood as due to quantum corrections, which can be studied in a controlled manner in terms of the parameters of the coherent states.
These parameters are the spins ${J_l}$, the classical extrinsic geometries ${K_l}$ and the spread $\sigma$, as well as the graph $\Gamma$ they are defined on. The interpretation of the semi-classical limit in terms of these parameters is rather subtle here.

%Besides the check that semiclassical states approximate the classical geometry they are peaking on, the actual aim is to analyze quantum corrections. For coherent states on a given complex these are parametrized by two parameters, i.e. the spins $\{J_{l}\}$ peaked on as well as the spread $\sigma$. Since the semiclassical limit is for large spins, quantum effects are expected to be stronger for small spins.

As far as the spread is concerned, with respect to the whole coherent state $\{J_{l},K_{l}\}$ there is a value where Heisenberg inequalities are minimized, which would be the `most semi-classical' case. On the other hand, since the spectral dimension on the spatial state is not dependent on the extrinsic curvature, obviously the deviation from the classical spectral dimension vanishes for $\sigma=0$, i.e., states sharply peaked at the intrinsic curvature but totally random in the extrinsic one. Nevertheless, these states are highly quantum. More interesting for our purpose are quantum effects of those states randomizing the intrinsic curvature, that is states with large spread $\sigma$.

Furthermore, from the physical interpretation of the geometric spectra with corresponding eigenbasis in terms of the spin representations, the limit $J_l\ra\infty$ is often seen as another semi-classical limit. 

Finally, one should not forget the dependence of the states on the underlying graphs. At least with respect to the spectral dimension, classically we have already seen the important dependence on the combinatorial structure. At the level of kinematical states, this dependence is still poorly understood in the literature.

In the following, we will consider the dual 1-skeletons of the previous regular torus triangulations as spin-network graphs $\Gamma$. 
They are parametrized by the number of nodes $|\Gamma_0| \propto p^d$ and their spectral dimension converges to the topological dimension $d$ if they are large or fine enough, i.e., in the limit $p \ra \infty$ (see section \ref{sub:Triangulations}).
Therefore, in such a geometric regime we can interpret the quantum corrections as actual deviations from the topological dimension.
Accordingly with this regular combinatorial structure, we will consider states peaked at all equal $J_l=J$ for $l\in \Gamma_1$. 

%\

A direct implementation of equation (\ref{eq:qds}) is unfeasible. This is because the number of terms in the quantum sum over representations $\{j_{l}\}$ even with cutoffs $j_{\rm min}$ and
$j_{\rm max}$ 
\begin{equation}
(j_{\rm max}-j_{\rm min})^{L}
\end{equation}
grows exponentially fast with the number of links $L=|\Gamma_{1}|=N_{d-1}$.
We have already shown that only for large classical triangulations (e.g., for $T^{2}$ of the order $10^{3}$; see figure \ref{fig:dsT2triang}) a geometric regime is obtained. 
Furthermore, although the state amplitude for many spin configurations vanishes due to the Clebsch--Gordan conditions implicit in the intertwiners, the resulting effective space of spin configurations is highly non-trivial and, in general, not well understood.

Alternatively, the quantum sum can be approximated by summing over
some number of configuration samples $\{j_{l}\}$ chosen randomly
according to the coefficients $\left|\psi(s)\right|^{2}$ (where the
norm must be included now to give a proper probability density).

If the space of representations is discrete, as for the spins of $SU(2)$, 
this measure needs to be discrete. Here we can choose the binomial distribution $B$. For large enough $\{J_{l}\}$ and $\sigma$, this is in turn well approximated by the the normal
distribution $\mathcal{N}$:
\begin{equation}
B\left(\left\lfloor\frac{2J^{2}}{J-\sigma^{2}}\right\rfloor,\frac{J-\sigma^{2}}{J}\right)\simeq\mathcal{N}\left(J,\frac{\sigma}{\sqrt{2}}\right)\,.\label{eq:BinomDis}
\end{equation}
The floor function $\lfloor\cdot\rfloor$ is needed since the first argument must be an integer. The variable $X=B(J,\sigma)$ is a random field dependent on $J$ and $\sigma$. The probability density function associated with it is, by virtue of the approximation (\ref{eq:BinomDis}), the Gaussian profile $(\pi\sigma^2)^{-1/2}\,\rme^{-(J-x)^2/\sigma^2}$.

Using this method, we have evaluated coherent spin networks on the $T^{2}$ triangulations discussed in section \ref{sub:Triangulations}.

To compare $\ds$ for various peaks $J$, we choose the scale $l_{*}$
such that $l^{2}(J)$ (the scale of the Laplacian) is kept fixed.
That is, we include a rescaling factor $r(J)=l^{2}(J)$: 
\begin{equation}
\left\langle \widehat{P(\tau)}\right\rangle _{\Gamma}^{\{J_{l},K_{l}\}}\propto\underset{\{j_{l}\}}{\sum}\left[\prod_{l\in\Gamma}\rme^{-\frac{(J_{l}-j_{l})^{2}}{\sigma^{2}}}\right]\mbox{Tr}\,\rme^{\tau r(J)\Delta_{\Gamma}(j_{l})}.
\end{equation}
Without this rescaling, as in the classical cases, one observes a shift $\ln\tau\to\ln\tau-\ln r(J)$ of the $\ds$ plot due to the $\Delta\to\Delta/r(J)$ scaling of the Laplacian. 
We will include this rescaling factor in the following calculations, mainly to allow for a more direct comparison of the spectral dimension of states.

\

\begin{figure}
\centering
\includegraphics[width=7.5cm]{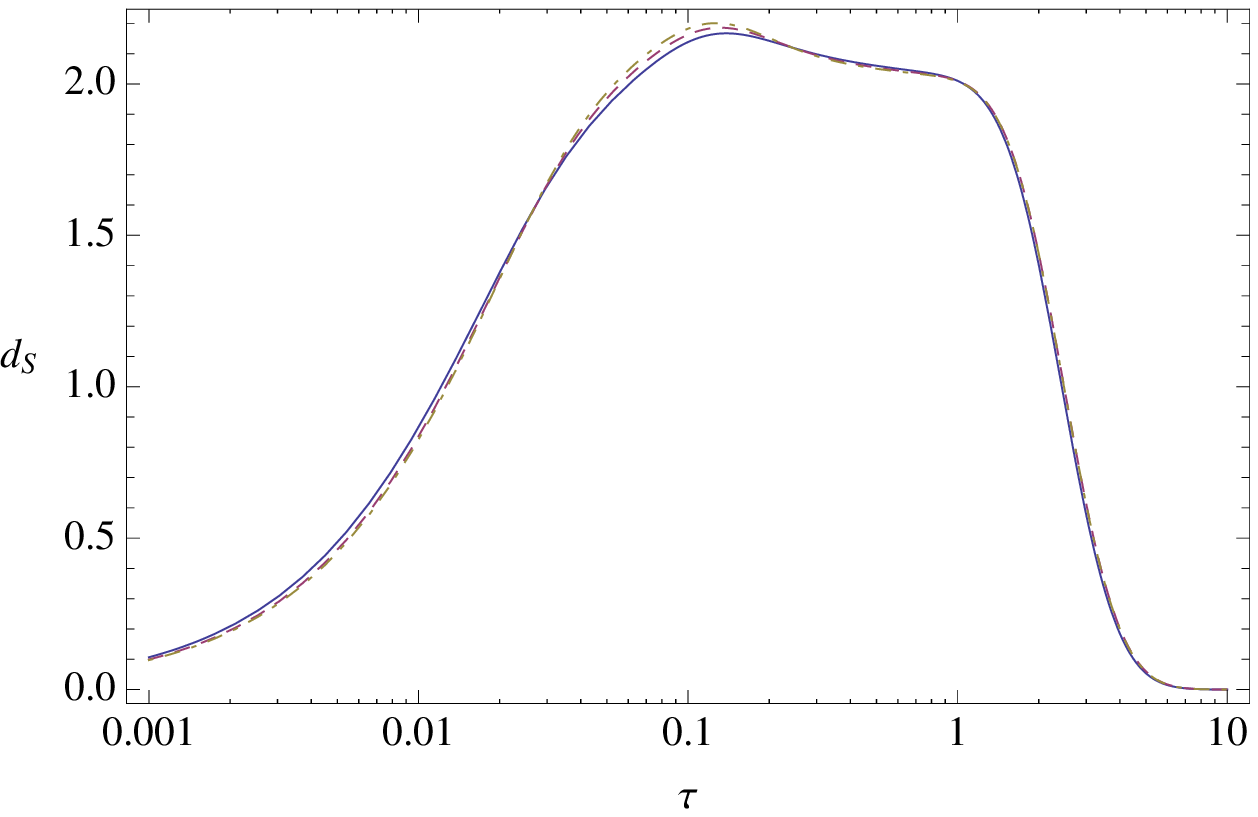}
\includegraphics[width=7.5cm]{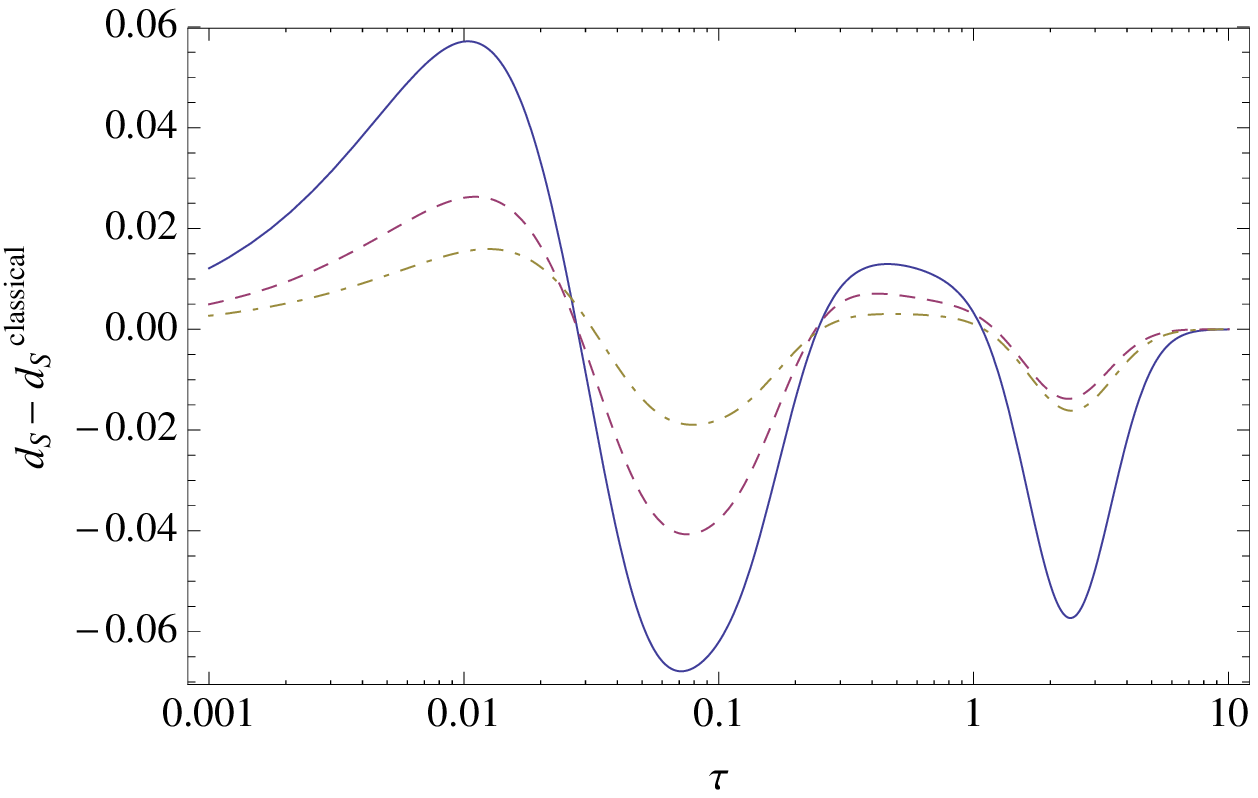}
\caption{Left: $\ds$ for coherent states peaked at $l(J)=J+1/2=16,32,64$ (solid, dashed and dot-dashed curve) on a regular torus triangulation (with $N_{2}=18\times4^{2}=288$ triangles) with spread $\sigma=\sqrt{J-1/2}$. Right: deviation of $\ds$ from the classical case.
\label{fig:coherJs}}
\end{figure}

We start with testing the dependence of coherent states with respect to the spin $J$ peaked at.
As expected, the quantum spectral dimension function of these states does not differ much from the classical version: 
for $J_{l}=J=\mathcal{O}(10)$ and $\sigma=\mathcal{O}(1)$, the deviation is at most of order $10^{-2}$ (figure \ref{fig:coherJs}). This is an important consistency check. 
Note, though, that all the known coherent states are peaked at discrete geometries. Therefore, strictly speaking, it is not the topological dimension but the particular spectral-dimension profile of these discrete geometries to be approximated well by the coherent states. What can be considered as quantum corrections in the spectral dimension is thus the difference between the quantum spectral dimension and the spectral dimension of discrete geometries.

\begin{figure}
\centering
\includegraphics[width=7.5cm]{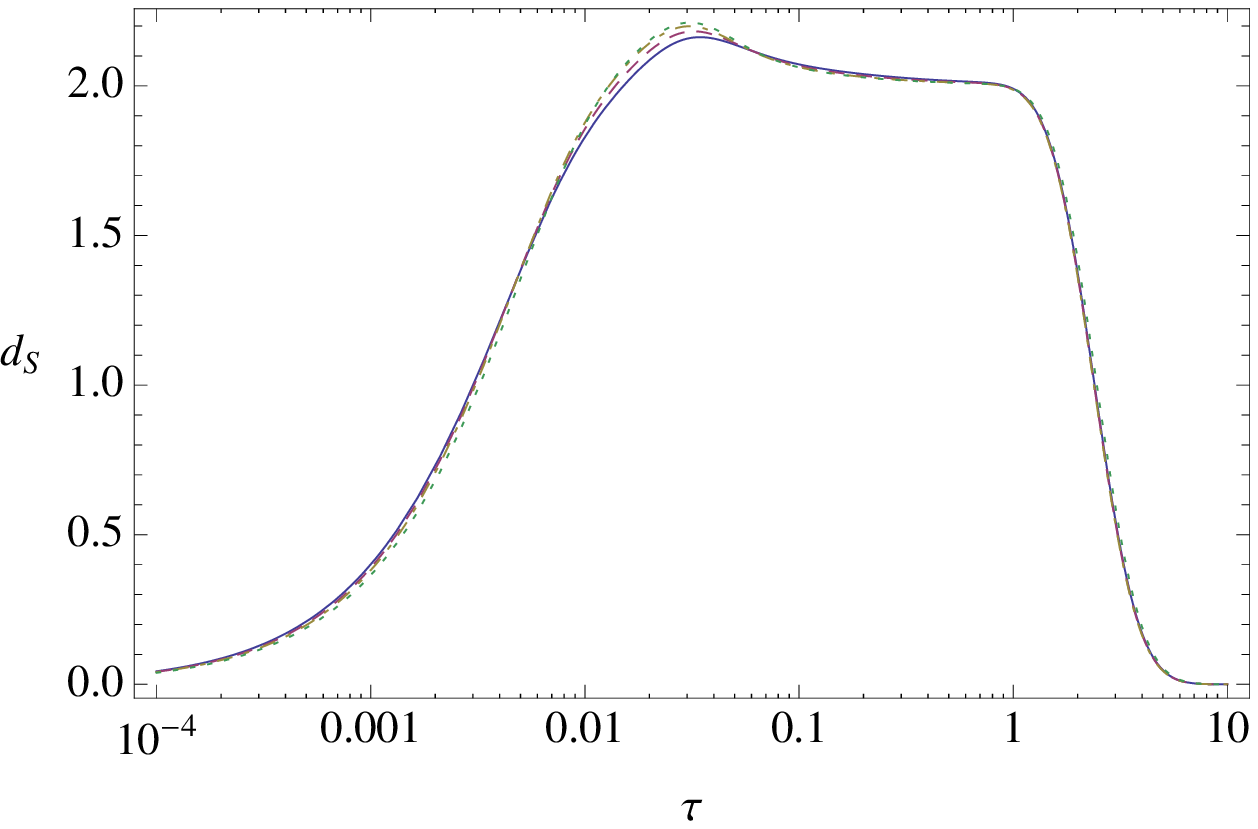}
\includegraphics[width=7.5cm]{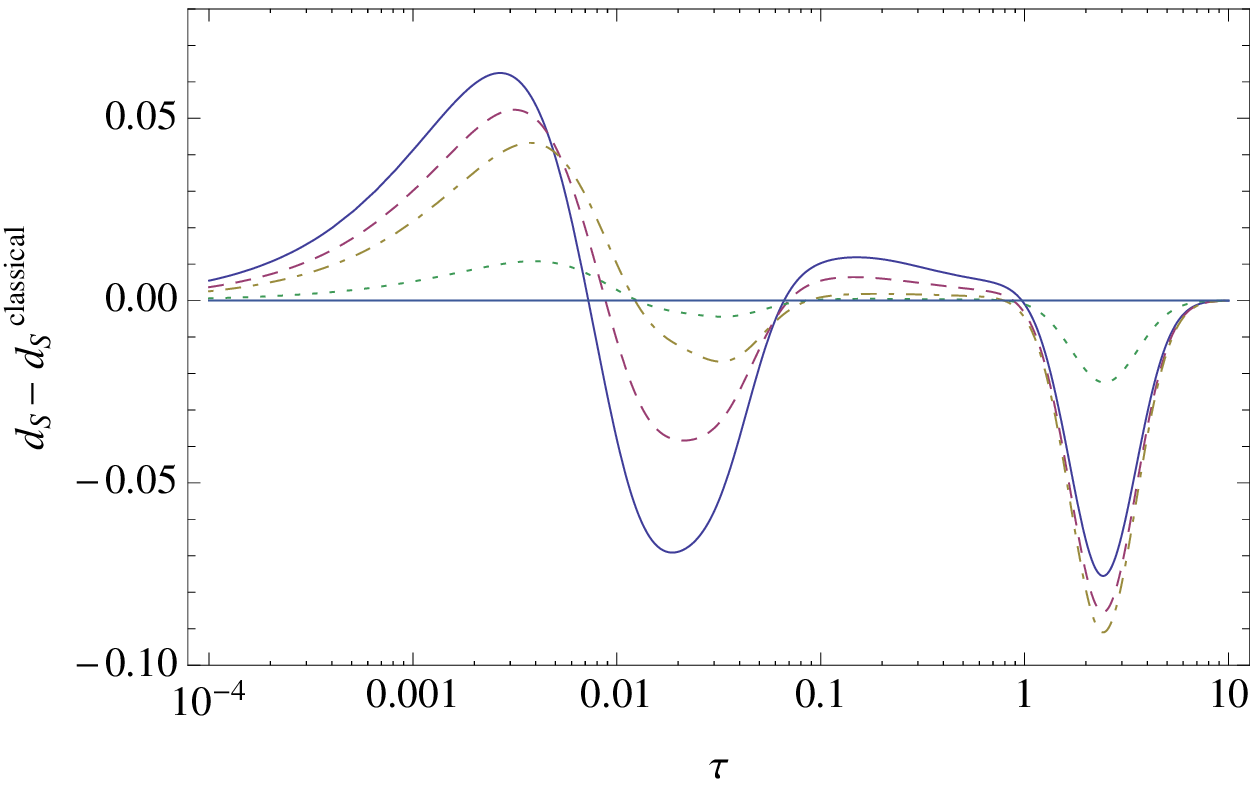}
\caption{Left: $\ds$ for coherent states peaked at $l(J)=J+1/2=16$ with spread $\sigma=1,2,3,\sqrt{15}$
(dotted, dot-dashed, dashed and solid curve) on a regular torus triangulation (with $N_{2}=18\times4^{3} = 1152$ right triangles).
Right: deviation of $\ds$ from the classical case.\label{fig:coherSigmas}}
\end{figure}

Second, we test the dependence of states with spread $\sigma$. Since the binomial distribution approximation (\ref{eq:BinomDis}) is only defined for $\sigma^{2}<2J$, it is difficult to probe the regime of larger $\sigma$ within this method. Probing the $\sigma$ dependence for $J_{l}=10$, we observe increasing quantum corrections, but still of order $\mathcal{O}(10^{-2})$ (figure \ref{fig:coherSigmas}).

%\

The main challenge when extending the calculations to larger spreads is
to deal with the highly non-trivial space of group representations due to
the Clebsch--Gordan conditions, as we noted. However, there is a very straightforward way
to define pure states (or their superpositions) by bounding the
range of spins to an interval $I=[j_{\rm min},j_{\rm max}]$ such that the
Clebsch--Gordan conditions are trivially fulfilled for any combinatorial (i.e., simplicial complex) structure of the states. In terms of the difference $J=j_{\rm max}-j_{\rm min}$, one could for instance construct states uniformly randomized over the interval
\begin{equation}
I_{J}=[\tfrac{1}{3}(2J+1),\tfrac{1}{3}(4J-1)]\cap\tfrac{1}{2}\N\,,
\end{equation}
on which any three elements satisfy the Clebsch--Gordan conditions.

\begin{figure}
\includegraphics[width=7.5cm]{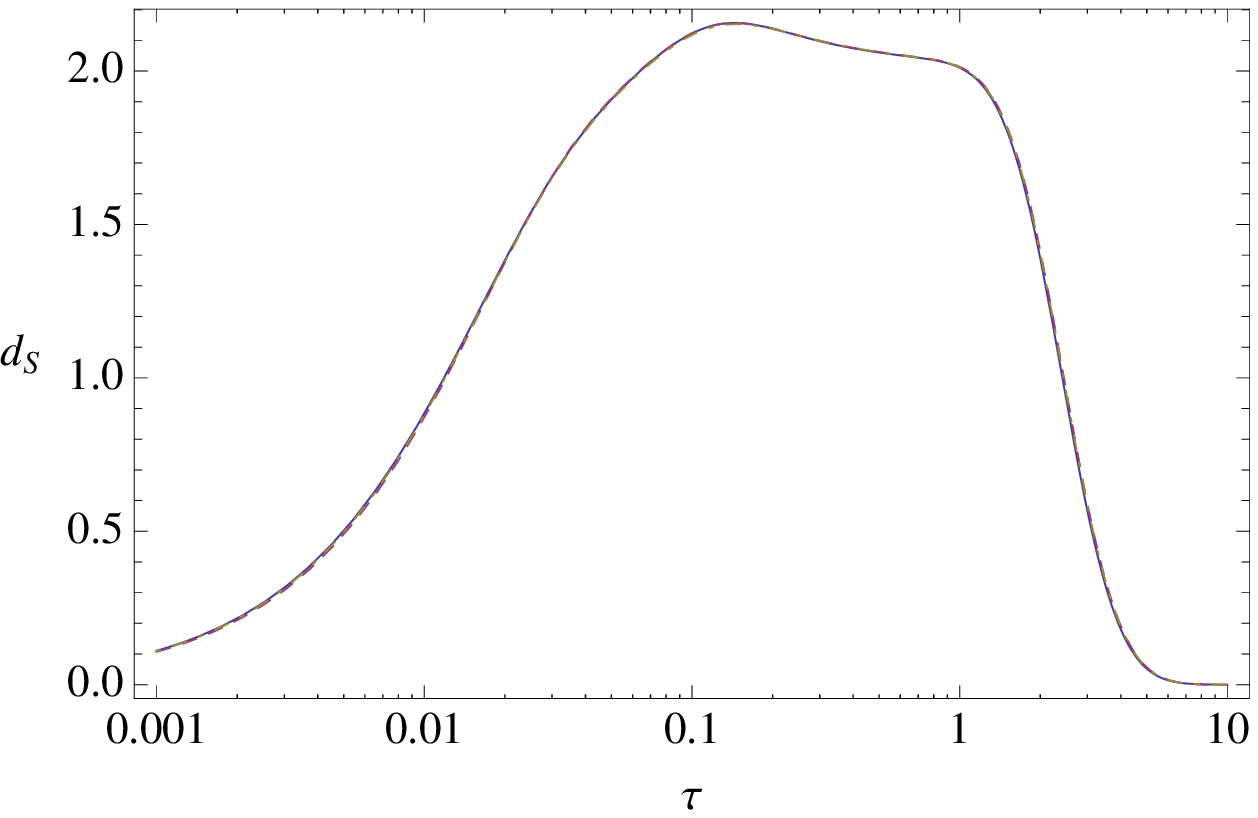}
\includegraphics[width=7.5cm]{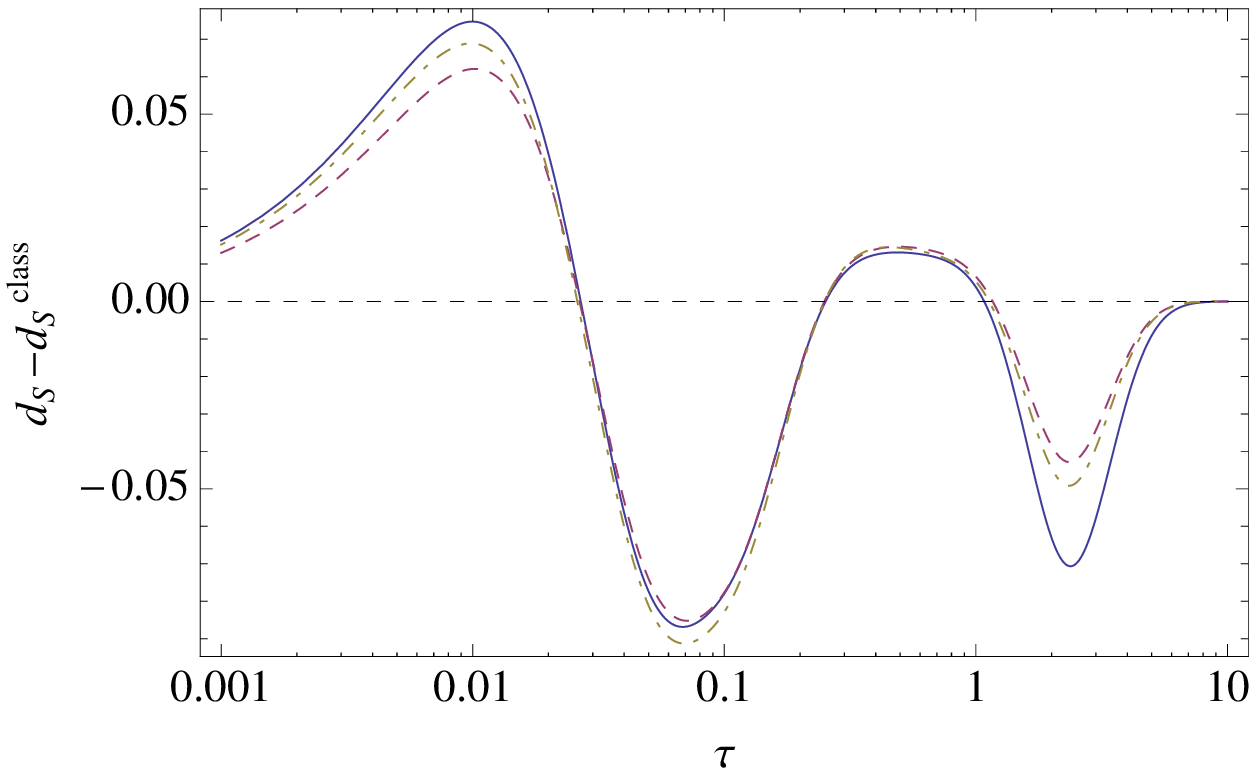}
\caption{Left: $\ds$ for a sum over seven random states with spins in $I_J$ for $J=15.5, 31.5, 63.5$ (solid, dashed and dot-dashed curve)  on a regular torus triangulation (with $N_{2}=288)$. Right: deviation of $\ds$ from the classical case.}
\label{fig:dsiid}
\end{figure}

As an example, we consider uniformly distributed spins for the same $J$ as in the above cases of coherent states (figure \ref{fig:dsiid}).
Again, the difference with respect to the classical triangulation randomized around is of order $\mathcal{O}(10^{-2})$.
It is worth noticing that, in all the above examples of quantum states, the difference with the corresponding classical state is rather marginal. The key result of the calculations for coherent states on a given regular triangulation is that the quantum corrections are very small, since by varying spins $J$ and spread $\sigma$ one gets deviations from the classical spectral dimension only of order $\sim 10^{-2}\ds$.

\subsection{Summing semi-classical (kinematical) states}

A basic feature of quantum mechanics is the superposition principle. The coherent and randomized states already are a typical example in this sense, as they are superpositions in the spin-network basis.
One can probe the quantum features of spatial geometry also by constructing other kinds of superposition states.
One obvious strategy is to superpose coherent states themselves. There are various ways one could do so.
Possible choices would be sums in the coherent-state labels $J$, $K$ and $\sigma$,
\begin{equation}\label{sup1}
|\Gamma,\{c_{J_{l},K_{l},\sigma}\}\rangle=\sum_{J_{l},K_{l},\sigma}c_{J_{l},K_{l},\sigma}|\psi_{\Gamma}^{\{J_{l},K_{l}\}}\rangle\,,
\end{equation}
or over different complexes and their dual graphs $\Gamma$:
\begin{equation}
|\{c_{\Gamma}\},J_{l},K_{l},\sigma\rangle=\sum_{\Gamma}c_{\Gamma}|\psi_{\Gamma}^{\{J_{l},K_{l}\}}\rangle\,.
\end{equation}
We will first deal with superposed states on the same complex and consider superpositions of complexes in the next section.

In the case (\ref{sup1}), the expectation value of the heat trace does not simplify to a single sum over the expectation values of squared coefficients. Assuming trivial intertwiners,
\begin{align}
\left\langle \widehat{P(\tau)}\right\rangle _{\psi}  &=
\sum_{J_{l},K_{l},\sigma}\sum_{J'_{l},K_{l}',\sigma'}\underset{\{j_{l'}\}}{\sum}\underset{\{j_{l}\}}{\sum}\overline{c_{J_{l},K_{l},\sigma}\psi_{\Gamma}^{\{J_{l},K_{l}\}}(j_{l})}c_{J'_{l},K_{l}',\sigma'}\psi_{\Gamma}^{\{J'_{l'},K'_{l'}\}}(j_{l}')\langle \Gamma,j_{l}|\mbox{Tr}\,\rme^{\tau\widehat{\Delta}}|\Gamma,j_{l}'\rangle\nonumber \\
 &= \sum_{J_{l},K_{l},\sigma}\sum_{J'_{l},K_{l}',\sigma'}\overline{c_{J_{l},K_{l},\sigma}}c_{J'_{l},K_{l}',\sigma'}\underset{\{j_{l}\}}{\sum}\prod_{l\in\Gamma}\overline{\rme^{-\frac{(J_{l}-j_{l})^{2}}{2\sigma^{2}}+\rmi K_{l}j_{l}}}\rme^{-\frac{(J'_{l}-j_{l})^{2}}{2\sigma'^{2}}+\rmi K'_{l}j_{l}}\mbox{Tr}\,\rme^{\tau\Delta_{\Gamma}(j_{l})}\nonumber\\
 &= \sum_{J_{l},K_{l},\sigma}\sum_{J'_{l},K_{l}',\sigma'}\overline{c_{J_{l},K_{l},\sigma}}c_{J'_{l},K_{l}',\sigma'}\underset{\{j_{l}\}}{\sum}\prod_{l\in\Gamma}\rme^{-\frac{1}{2}\left[\frac{(J_{l}-j_{l})^{2}}{\sigma^{2}}+\frac{(J'_{l}-j_{l})^{2}}{\sigma'^{2}}\right]-\rmi(K_{l}-K'_{l})j_{l}}\mbox{Tr}\,\rme^{\tau\Delta_{\Gamma}(j_{l})}.
\label{eq:htCrossterms}
\end{align}
Assuming a fixed extrinsic curvature $K_{l}$ there
is no phase causing interferences, and for sharply peaked coherent
states the cross terms with $J_{l}\ne J'_{l}$ are suppressed. In
this case, one would expect that the `naive' sum disregarding the cross terms
\begin{equation}
\sum_{J_{l},\sigma}\underset{\{j_{l}\}}{\sum}\left|c_{J_{l},\sigma}\right|^{2}\left[\prod_{l\in\Gamma}\rme^{-\frac{(J_{l}-j_{l})^{2}}{\sigma^{2}}}\right]\mbox{Tr}\,\rme^{\tau\Delta_{\Gamma}(j_{l})}
\end{equation}
would still be a good approximation for $\langle\hat P\rangle$.

\begin{figure}
\centering{}
\includegraphics[width=7.5cm]{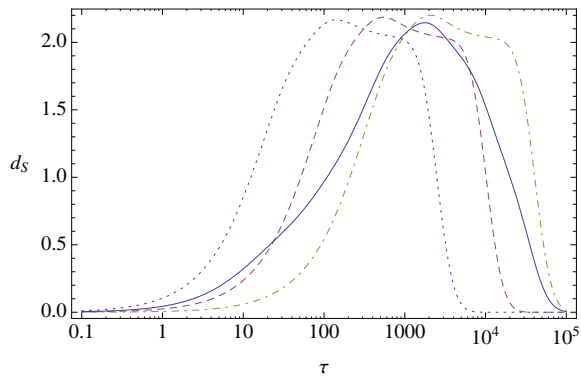}
\caption{$\ds$ for the superposition of coherent states $l(J)^2=J+1/2=16,32,64$ on a regular torus triangulation (with $N_{2}=18\times4^{2}=288$ triangles) with unrescaled Laplacian compared with the states summed over (dashed curve; see figure \ref{fig:coherJs}).}
\label{fig:SupJs}
\end{figure}

\begin{figure}
\includegraphics[width=7.5cm]{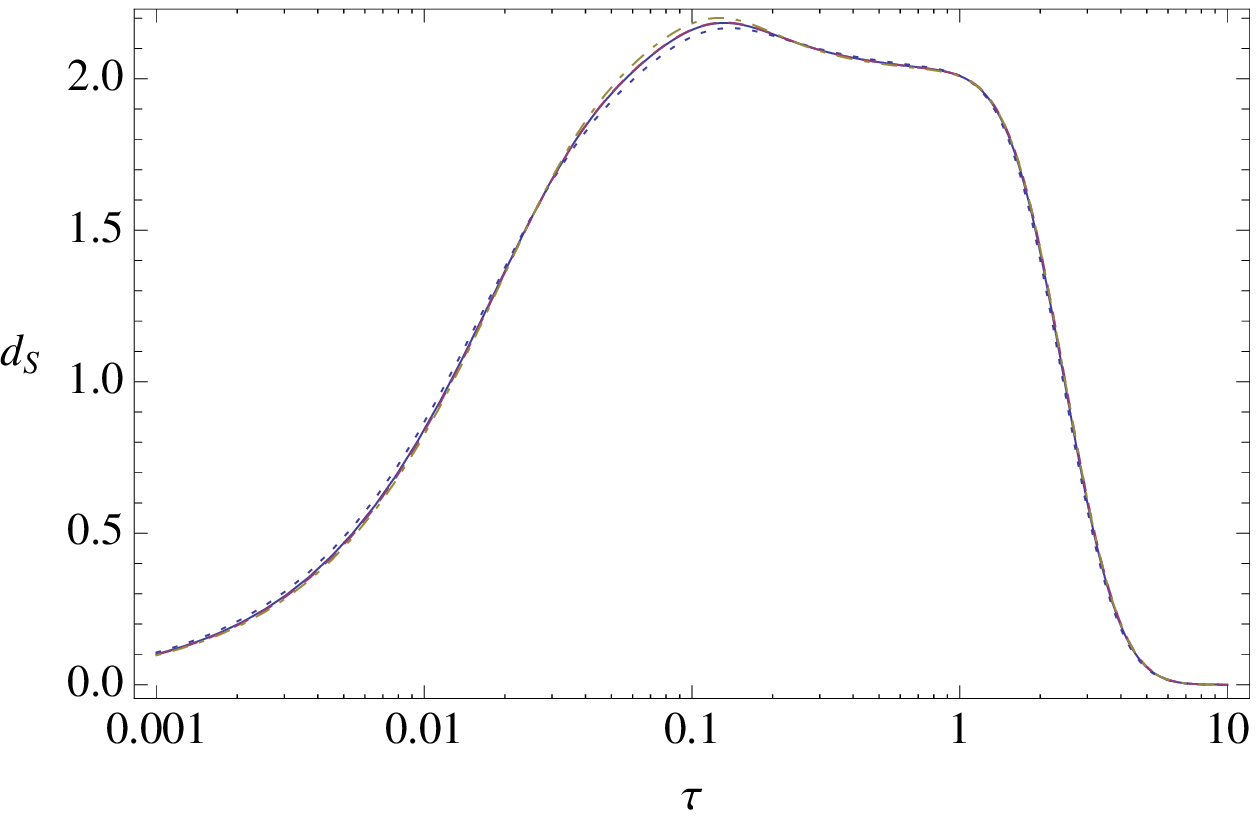}
\includegraphics[width=7.5cm]{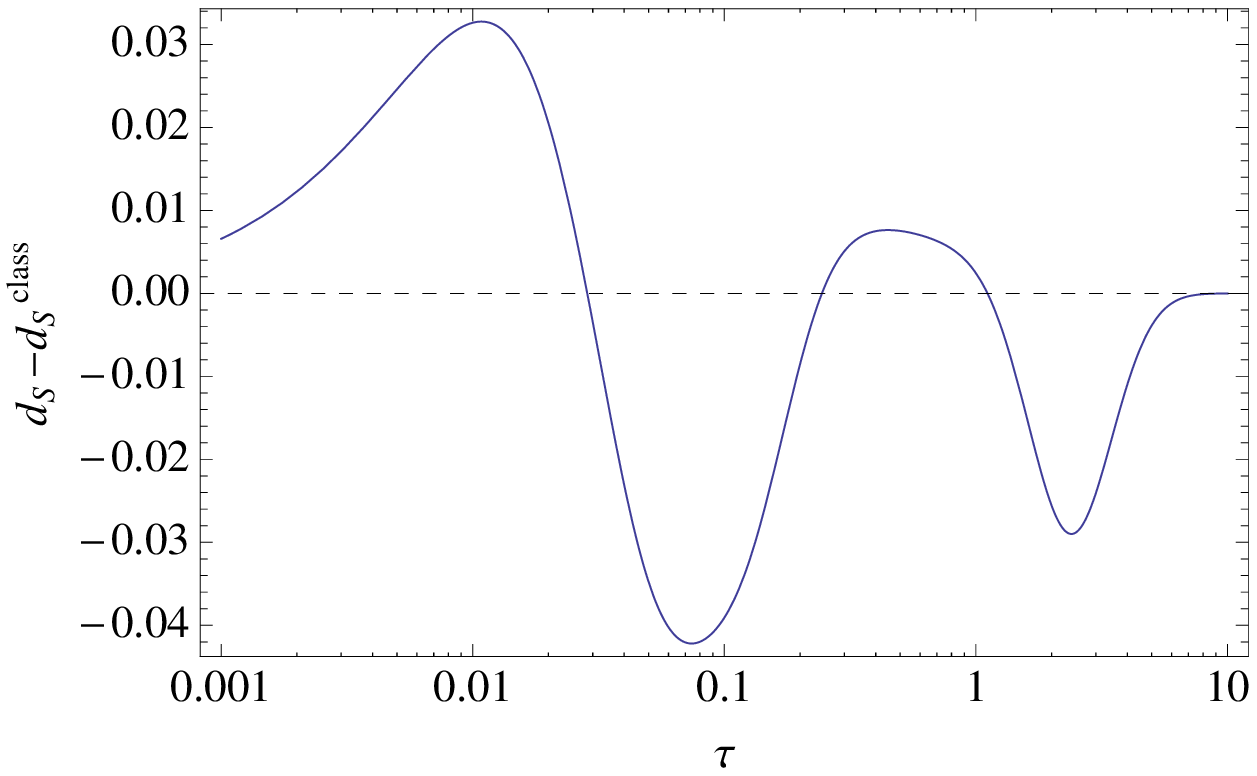}
\caption{Left: $\ds$ for the superposition of coherent states $l(J)^2=J+1/2=16,32,64$ on a regular torus triangulation (with $N_{2}=18\times4^{2}=288$ triangles) with rescaled Laplacian compared with the states summed over (dashed curve; see figure \ref{fig:coherJs}). Right: deviation of $\ds$ from the classical case.}
\label{fig:SupJref}
\end{figure}

When summing over coherent states peaked at different spins, it makes a crucial difference whether we take them as peaked at classical geometries of different size or at the same classical geometry given by a fixed scale $l = l_\star J$. 
In both cases, the spectral dimension turns out to be an average of the spectral dimensions of the parts summed over. However, in the first case (figure \ref{fig:SupJs}) these individual profiles are shifted with respect to one another such that the superposition has dimension of order one only in the regime where most of them overlap. 
In the second case (figure \ref{fig:SupJref}), all profiles have features at the same scales, so that the superposition yields a $\ds$ plot close to the classical geometry peaked at; the quantum correction is even less important, being smaller than for the individual coherent states (figure \ref{fig:coherJs}).
%of the order $\mathcal{O}(10^{-2})$.

\

In this section, we have investigated the spectral dimension of coherent states and superpositions of these on toroidal complexes. Quantum corrections are small in the small-spin regime (figure \ref{fig:coherJs}), for large spreads (figure \ref{fig:coherSigmas}) and for total randomization in a given spin interval (figure \ref{fig:dsiid}). Superpositions of coherent states peaked at the same geometry smoothen out the quantum deviations even more (figure \ref{fig:SupJref}). Only superpositions of states peaked at different geometries show a distinct behaviour. This happens when the spectral dimension of the individual states is noticeably different from one another (figure \ref{fig:SupJs}).

%%%%%%%%%%%%%%%%%%%%%%%%%%%%%%%%%%%%%%%%%%%%%%%%%%%%%%%%%%%%%%%%%%%%%%%%%%%%%%%%%%%%%%%%%%%%%%%%

\section{Summing complexes\label{sec:sumDis}}

In quantum gravity, another source of non-trivial features in effective geometry
is expected to come from summing over the combinatorial structures. 
Thus, in this section we will investigate superpositions of states on distinct combinatorial complexes\footnote{In CDT these are the only dynamical degrees of freedom. The fact that the spectral dimension of the spacetime sum-over-histories is scale dependent in the CDT ensemble \cite{Ambjorn:2005fj,Benedetti:2009bi} is thus a consequence of (and eventually needs to be explained by) the way it is summed over a class of simplicial manifolds.
Although here we restrict ourselves to kinematical spatial states and their superpositions, when summing with certain weights over simplicial manifolds we are in a setting quite comparable with CDT, and similar results could be expected. The main difference is that, while in CDT there is a precise description for the integration measure given by the exponential of the Regge action of equilateral triangulations, in the context of kinematical states of quantum geometry there is no unique prescription for how these states should be superposed on different complexes, in order to obtain some approximately smooth geometry. Therefore, our considerations are somewhat explorative and driven mainly by the aim of unveiling generic features of superpositions of complexes.}.

In the following, we are interested in superposed states of the form
\begin{equation}
\left\langle \widehat{P(\tau)}\right\rangle \propto\sum_{\Gamma}\underset{\{j_{l}\}}{\sum}w_{\Gamma}^{2}\left|\psi_{\Gamma}^{\{J_{l},K_{l}\}}\right|^{2}\langle \Gamma,j_{l}|\mbox{Tr}\,\rme^{\tau\widehat{\Delta}}|\Gamma,j_{l}\rangle .
\label{eq:SupGraphs}
\end{equation}

In this definition, we have included a weighting factor $w_{\Gamma}$ for the graphs to be superposed. A natural choice would be, for example, $w_{\Gamma}=1/\mbox{sym}\Gamma$, a symmetry factor for the automorphism group, which may or may not be included. The calculations we will present do not include it, but later on we will briefly discuss the effect of weighting, which we checked in detail. Comparing with the heat-trace expectation value of the superposition of coherent states on the same complex, equation (\ref{eq:htCrossterms}), the sum here does not contain cross terms since spin-network states on different graphs are orthogonal. 

%{\bf J: the interesting case of summing many (above 40) lattices seems not to be feasible; summing just a few is not as interesting. Only smaller deviations to the above case summing just equilateral triangulations would be expected, anyhow}  

We will only consider the summation over equilateral triangulations. These can be understood as the extreme case of minimal spread $\sigma$, that is, sharply peaked at the intrinsic curvature but totally randomized in the extrinsic one.
The reason is that we have already seen in the previous section that the effect of the quantum fluctuations in coherent states is only of order $10^{-2}$. One would then expect that the effect of superposing truly quantum coherent states can be reproduced by a superposition of the discrete geometries associated with their peak values (equation \Eq{eq:SupGraphs}). Indeed, we will find that the effects of superposing graphs are of order higher than $10^{-2}$.
This justifies the approximation of the full sum, which in the case of superposition of graphs is considerably more challenging from a numerical point of view.

%\

%\subsection{Superposition of classical lattices\label{sub:Summing-over-lattices}}

\begin{figure}
\centering{}
\includegraphics[width=7.5cm]{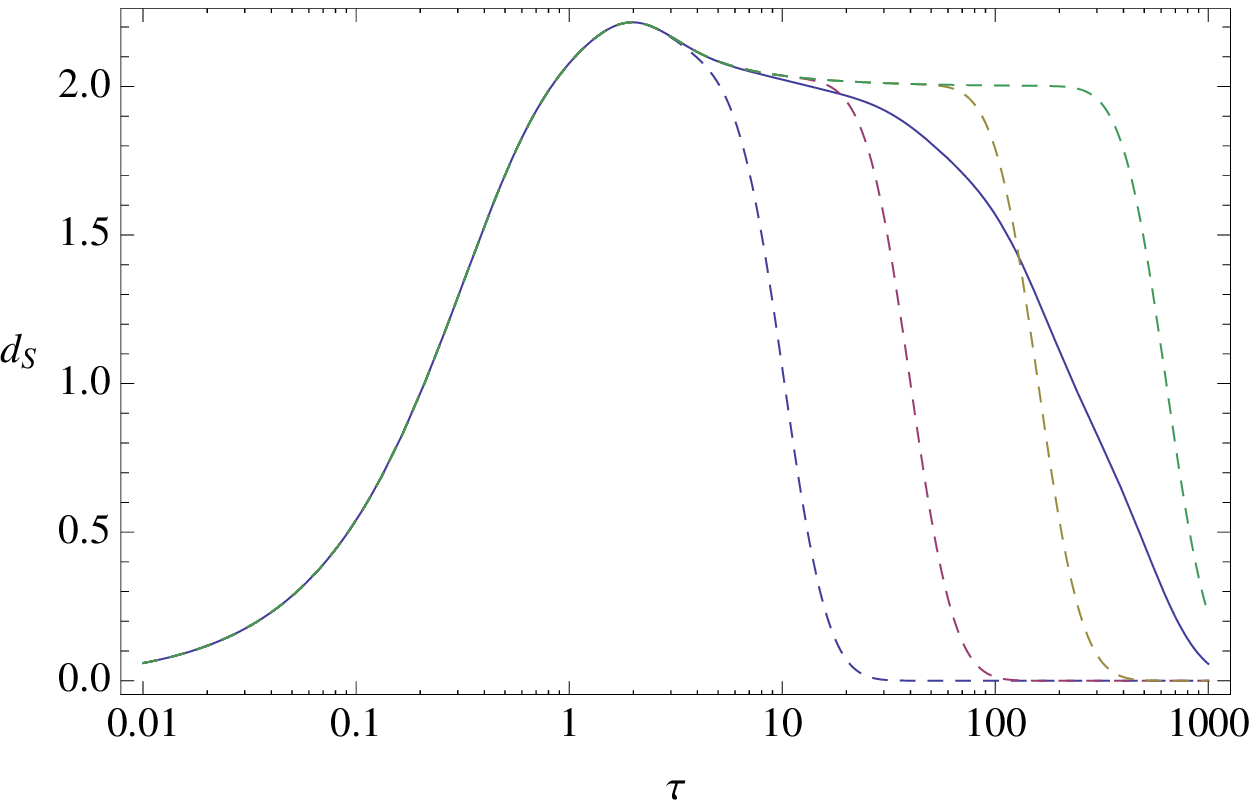}
\includegraphics[width=7.5cm]{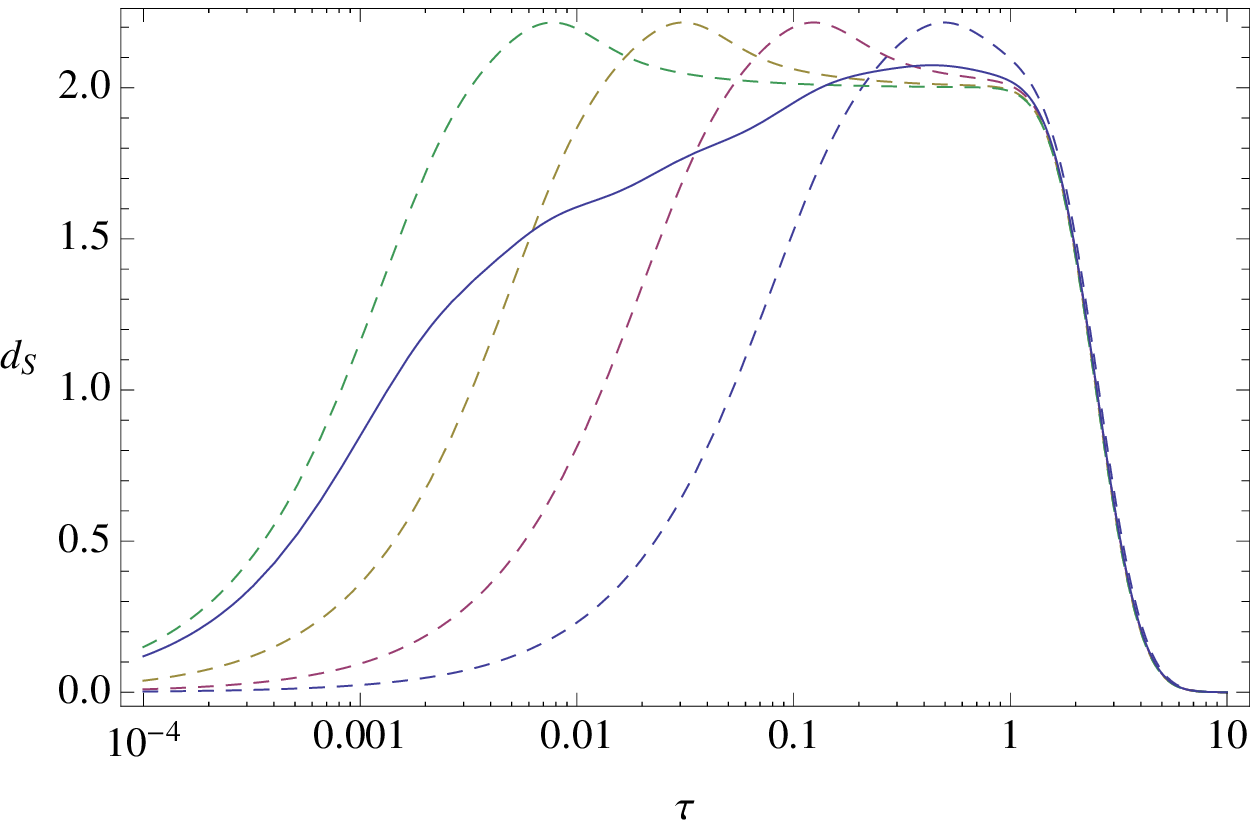}
\caption{Sum over the regular $T^{2}$ triangulations of figure \ref{fig:dsT2triang} (included as dashed curves for comparison) with trivial weights, unrescaled (left) and rescaled (right).
\label{fig:sumT2}}
\end{figure}

%We have already seen that the corrections to the spectral dimension of coherent states with respect to the classical discrete geometry at which they are peaked is only of the order of $10^{-2}$. It is therefore useful to consider first a superposition of the equilateral geometries. Namely, we calculate the spectral dimension from a superposition of some classical heat traces $P^{(k)}(\tau)$, possibly weighted by some factors $w_{k}$:
%\begin{equation}
%P(\tau)=\sum_{k=1}^{K}w_{k}P^{(k)}(\tau).
%\end{equation}
%This should be understood as an expectation value of a superposition state exactly analogue to eq. \ref{htSuperposition}.
%In particular, we are interested in the $P^{(k)}$ of the equilateral torus triangulations of different size considered in section \ref{sub:Triangulations}.
As before, it is crucial whether the superposed triangulations are taken as purely combinatorial or as refinements of the same smooth geometry through rescaling the edge lengths (figure \ref{fig:sumT2}). While in both cases the result is an averaging of the spectral dimension (depending on the weights $w_{\Gamma}$), in the first one this effect takes place at larger diffusion scales, in contrast with the small-diffusion-scale regime of the second.

The most interesting case here is the second (rescaling). Since all complexes summed over are triangulations of the same smooth geometry, it can be interpreted as a special case of a semi-classical state incorporating a (kinematical) continuum approximation (which, obviously, can only be implemented up to some finite order in our explicit calculations). Furthermore, the peak in the spectral dimension is a discreteness effect which does not appear for these states as a result of averaging.

\begin{figure}
\centering{}
\includegraphics[width=7.5cm]{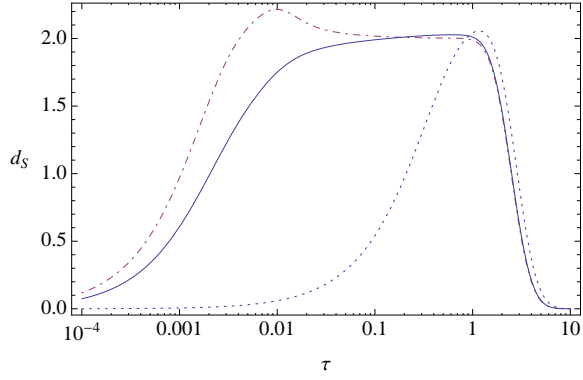}
\caption{Sum over all rescaled regular $T^{2}$ triangulations of size $p=3,4,\dots,43$ (solid curve) and, for comparison, the individual $p=3$ and $p=43$ cases (dotted and dot-dashed curve).
\label{fig:sumT2pmax}}
\end{figure}

For these reasons, it is interesting to calculate not only the superposition of a few rescaled triangulations but also of all regular triangulations of the type described in section \ref{sub:Triangulations}, of size ranging from $p > 3$ up to some $p_{\rm max}$ (figure \ref{fig:sumT2pmax}).
The effect is a more extended plateau. Thus, from these calculations one would expect the appearance of a plateau at the topological dimension for sums over more and larger triangulations, but without the discretization effect of a peak at the characteristic lattice scale.

Finally, the same can be done for the randomly subdivided triangulations. The effect of summing is, again, an averaging of the $\ds$ profiles. Qualitatively, these are quite different from the regular triangulations. %More precisely we recover the same behaviour of these classical triangulations already discussed before.
% A plateau could be conjectured somewhere around $d\approx 1.5$.
\begin{figure}
\centering{}
\includegraphics[width=7.5cm]{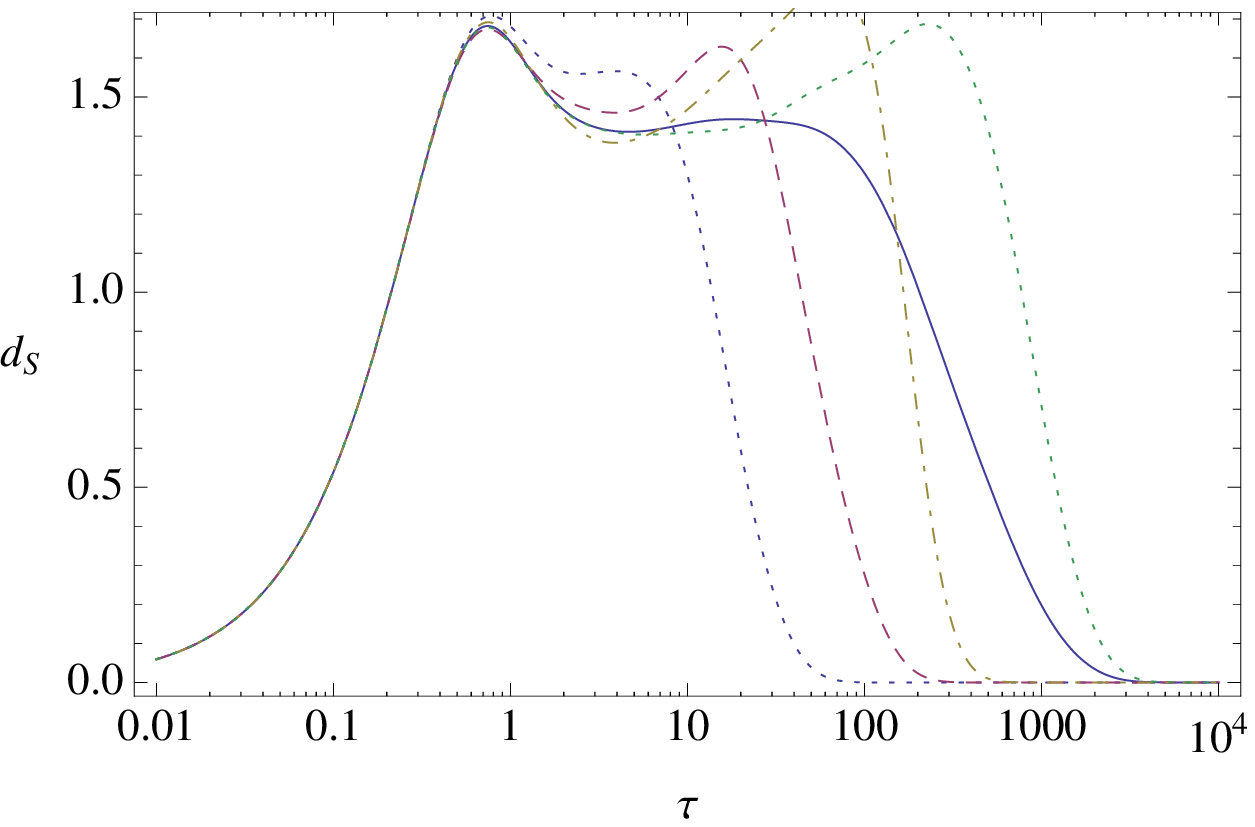}
\includegraphics[width=7.5cm]{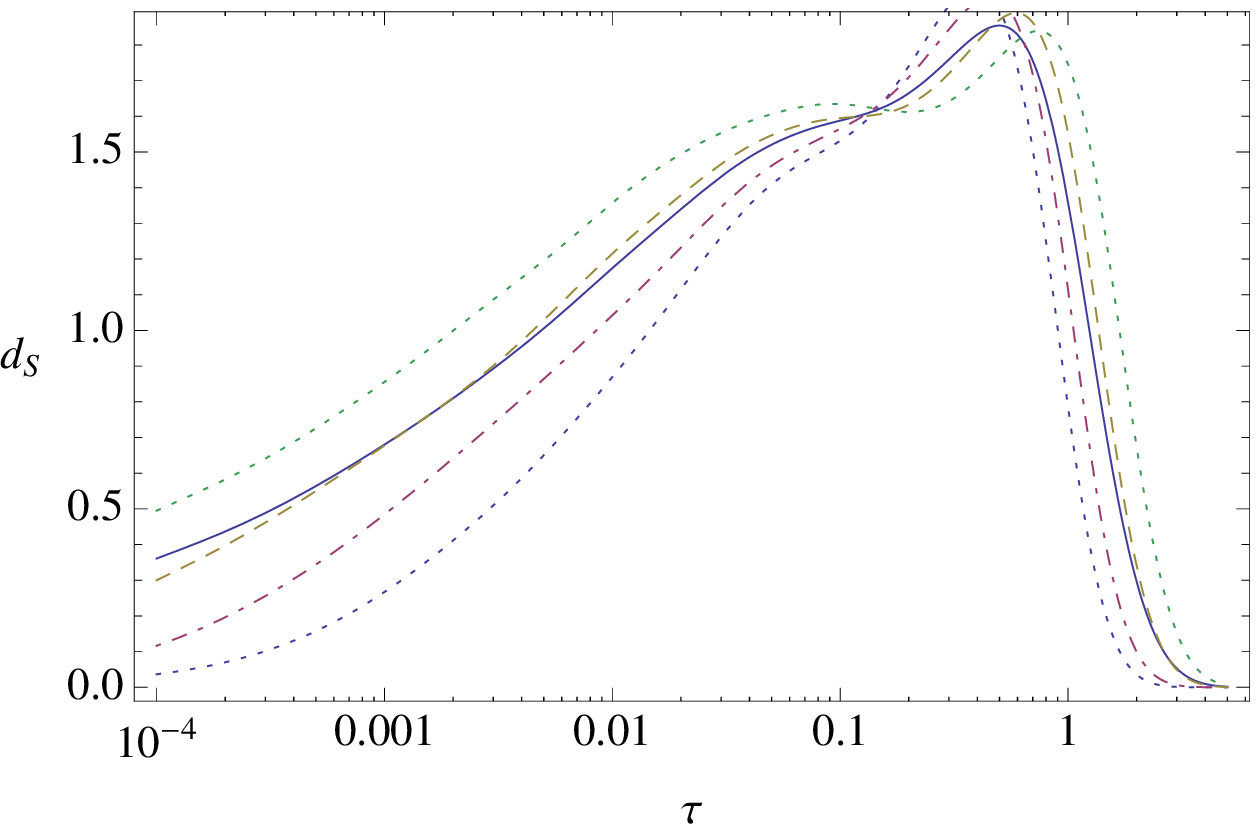}
\caption{Sum over randomly subdivided $T^{2}$ triangulations (dashed line, for comparison; see figure \ref{fig:dsT2subran}), unrescaled (left) and rescaled (right).
\label{fig:dsT2randomsum}}
\end{figure}

%\subsection{Superposition of coherent states on different triangulations}

\

In this final section, we have seen that the superposition of states on distinct combinatorial complexes produces an interesting behaviour of the spectral dimension. For a superposition of regular triangulations of the same smooth geometry but of different size, we find that the discreteness artifact of the the peak, appearing in classical cases, disappears (figure \ref{fig:sumT2}) and a plateau with the topological dimension is obtained (figure \ref{fig:sumT2pmax}). Still, there are no hints for an effective smaller dimension at smaller scales in one superposition state. The running of the spectral dimension in the profiles of the figures is, in fact, mainly due to discreteness artefacts ($\ds\to 0$ at small $\tau$) and topological effects ($\ds\to 0$ at large $\tau$). On the other hand, superpositions of 1-3 Pachner subdivisions have a $\ds$ plateau at a height smaller than the topological dimension (figure \ref{fig:dsT2randomsum}). Summarizing, the averaging effect stemming from superposition states is the only manifestation of additional geometric data in the profile of the spectral dimension, which would be otherwise reproduced by a classical triangulation.

%%%%%%%%%%%%%%%%%%%%%%%%%%%%%%%%%%%%%%%%%%%%%%%%%%%%%%%%%%%%%%%%%%%%%%%%%%%%%%%%%%%%%%%%%%%%%%%%

\section{Summary and outlook}

In this paper, we have explored the spatial spectral dimension of states of quantum geometry in the case of (2+1)-dimensional Euclidean LQG. To this end, we have discussed the definition of the spectral dimension for combinatorial complexes equipped with the geometric variables defining the LQG configuration space. We have then showed that quantum states have a well-defined spectral dimension, identified with the scaling of the quantum expectation value of the heat trace. We also checked the quantum observable properties of these LQG geometries.

We have presented for the first time a systematic classification of topological effects and discreteness artefacts in classical settings, pushing the analysis to analytic expressions whenever possible. After that, we studied coherent states in LQG, both as superpositions of geometries on the same complex and as superpositions of complexes, in particular triangulations of the same geometry. One of the main results of this work is the lack of any strong indication of a dimensional flow \emph{not} due to topology or discreteness, as otherwise suggested in \cite{Modesto:2009bc}. 

To clarify the source of this discrepancy, let us discuss the differences between our work and \cite{Modesto:2009bc} more in detail. The basic idea in \cite{Modesto:2009bc} is to consider the scaling of the (quantum) Laplacian with a length scale $\ell$.  Since the Laplacian scales as the inverse metric, and the metric roughly scales as an area, one could relate the scaling of the spectrum of the area operator in LQG to the scaling of the quantum Laplacian. With the further identification, on one hand, of the length scale with the spin representation label $j$ as $\ell = l_{\rm Pl}\sqrt{j}$, and from dimensional reasonings in momentum space on the other hand, one ends up with an effective dispersion relation which is eventually used to derive the spectral dimension, as in the smooth approaches mentioned in the introduction.

Our starting point, in contrast, is to consider the full structure of the Laplacian as directly acting on states of quantum geometry. The proper discrete Laplacian (\ref{eq:disLap}) indeed turns out to have a much richer structure than in the inverse-area heuristic argument. It is not obvious whether there could be any regime of approximation where the dimensional reasoning about the inverse area would apply.

The Laplacian takes necessarily the form of a matrix with non-zero entries for incident, neighboring nodes. Therefore, it is further required to take the full underlying combinatorial structure into account, which is neglected in the approximations of \cite{Modesto:2009bc}. Also, the validity of an approximation based on a single simplex is hard to understand from our perspective, as we have found that it is crucial to consider large enough complexes in order to reproduce the topological dimension at large scales (small probing resolution). As we have seen, calculations on triangulations with few simplexes, such as the dipole, do not show any geometric regime between the ones in which discreteness and topology effects dominate, thus forbidding any interpretation of the spectral dimension as a spatial dimension.

Nevertheless, one might try to understand the approximations of \cite{Modesto:2009bc} in terms of regular lattices such as the ones used here with the regular triangulations. However, in contrast with the non-trivial dispersion relation derived from the area spectrum there, in our setting the effect of the discrete (length) spectrum seems to be washed away in a twofold way. First, because of the more involved manner in which the spectra are entering the expression of the Laplacian and, second, because of the tracing over the dual nodes of large complexes. These two differences seem to be the main reason for the distinct results. Due to the complexity of the problem, we have been unable to obtain the explicit dependence on such geometric data in the spin representation (pure or coherently peaked at) in a closed, analytic form. Yet, this can still be analyzed when comparing states labelled with different spins. For the latter, our results clearly show only a minimal dependence (figure \ref{fig:coherJs}). 

Notwithstanding, it has to be stressed that the calculations we have presented are only a first step, so our results can only be seen as a hopefully interesting indication of how actual LQG geometries behave. First, we have analyzed the spectral dimension only for \emph{kinematical} states and in 2+1 Euclidean LQG. In \cite{Modesto:2009bc}, the argument was for kinematical states in 3+1 dimensions, although the result was interpreted as the spectral dimension of the full (3+1)-dimensional spacetime and the effective dispersion relation was inserted in a heat trace over four-dimensional momentum space. Furthermore, the argument was extended to physical states on certain one-simplex spin foams in the boundary formalism, in 2+1 dimensions \cite{Caravelli:2009td} as well as in 3+1 dimensions \cite{Magliaro:2009wa}. The spectrum yielding the effective dispersion relation was taken to be the same in both dimensions there, but for the full Laplacian considered in the present paper the step to 3+1 dimensions gives rise to essential differences, as already explained in the introduction. It is also clear that, eventually, the quantum spectral dimension should be considered on physical states, i.e., the quantum dynamics needs to be taken into account for the results to have a solid physical significance. Let us briefly discuss how these issues could be addressed.

%\

The most significant differences for gravity in 3+1 dimensions versus 2+1 dimensions lie in the dynamics. In this respect, one would not necessarily expect crucial qualitative discrepancies at the level of kinematical states in the quantum theory. However, in LQG the polyhedral interpretation in terms of the spectra of geometric observables is different in that the length spectra in 2+1 dimensions already provide a classical geometric interpretation as Regge geometries, while in 3+1 dimensions the corresponding geometric variables are not sufficient. More variables such as dihedral angles would be needed, but their corresponding quantum operators do not commute.

In principle, these uncertainties do not pose any obstacle to define the Laplacian \cite{COT1} and analyze the quantum spectral dimension. Such features could be effectively captured using the flux representation of states and operators \cite{Baratin:2011hc} or, alternatively, the spinor representation \cite{Livine:2011uh}. Beside the implicit operator ordering ambiguities, the major challenge in pursuing this step is just technical. The explicit form of the Laplacian becomes very complicated and clever ideas and approximations would be needed to cast the expectation value of the corresponding heat trace into a form feasible for actual computations. This is going to be an interesting challenge, as the non-commutativity in 3+1 dimension might give rise to unexpected features in the quantum spectral dimension.

%\

From the point of view of a complete theory of quantum gravity, the challenge to include the full dynamics is more important.
%This is a potential source of new characteristics of the spectral dimension on spatial states when taking the expectation value of the heat trace in terms of a physical inner product.
This is indeed the starting point for studying the spectral dimension of spacetime, which is the actual quantity mainly studied in the quantum gravity literature so far.
This observable could be defined in a covariant, path-integral-like formalism. Then, the expectation value of the heat trace would amount to the insertion of the heat trace function in appropriate variables into a sum over spacetime histories. Independently of the precise model of LQG dynamics chosen (i.e., the exact spin-foam or, equivalently, group field theory model), new issues arise.

A first challenge is the concept of spacetime heat trace as a path-integral observable. The latter is not simply a polynomial in the configuration variables. This raises several computational difficulties: the insertion of the heat trace in the path integral cannot be easily split into integrals of monomials of local variables. As a result, it seems quite hard, if not impossible, to solve the path integral with such heat-trace insertion. One should then find some viable approximation, for example one in which earlier results \cite{Livine:2009bz} on `ultralocal' observables (i.e., observables depending only on the moduli of the discrete $B$-field values) in the Ponzano--Regge model can be used to evaluate the heat-trace expectation value.

Even if a method for solving the path integral was at hand, a further issue is raised by degenerate configurations in the state sum. Already in the (2+1)-dimensional theory, the spin-foam sum generically contains degenerate geometries, in contrast with its kinematical states as shown in section \ref{sec:defLQG}. We have done preliminary calculations on small complexes showing this feature \cite{COT3}. As a consequence, the spacetime heat-trace insertion has a complex part. This is not, in itself, a sign of any inconsistency. Nevertheless, it raises an interesting question about the physical relevance and interpretation of a complex spectral dimension, and it may represent an intriguing line of investigation in the analysis of the spin-foam quantum dynamics.

Finally, a general issue with any explicit spin-foam evaluation is divergences. This is crucial for a global object such as the spectral dimension, since common regularization techniques may not be appropriate. For instance, gauge-fixing geometric data along a maximal tree of the dual complex \cite{Freidel:2004bq} are problematic because the heat trace might trivialize on the corresponding configurations. Considering coherent states on the boundary suppressing larger spin configurations (the cause of divergences) in the bulk works only for smaller complexes. But, as we have discussed, larger complexes are needed in order to obtain a meaningful spectral dimension. One way to circumvent this problem would be spin-foam models with quantum groups \cite{han}. An actual solution, however, would consist in the proper renormalization of spin foams in terms of their group field theory (second-quantized) description \cite{Carrozza:2012th,Carrozza:2013vz}. 

%%%%%%%%%%%%%%%%%%%%%%%%%%%%%%%%%%%%%%%%%%%%%%%%%%%%%%%%%%%%%%%%%%%%%%%
%%%%%%%%%%%%%%%%%%%%%%%%%%%%%%%%%%%%%%%%%%%%%%%%%%%%%%%%%%%%%%%%%%%%%%%

\ack{The work of GC is under a Ram\'on y Cajal contract. DO acknowledges financial support from the A. von Humboldt Stiftung, through a Sofia Kovalevskaja Award. JT acknowledges financial support by the Andrea von Braun Foundation}

%%%%%%%%%%%%%%%%%%%%%%%%%%%%%%%%%%%%%%%%%%%%%%%%%%%%%%%%%%%%%%%%%%%%%%%
%%%%%%%%%%%%%%%%%%%%%%%%%%%%%%%%%%%%%%%%%%%%%%%%%%%%%%%%%%%%%%%%%%%%%%%

\

\providecommand{\href}[2]{#2}
\begingroup
\raggedright
\endgroup

\end{document}